\documentclass[aps,pra,reprint]{revtex4-1}
\usepackage[latin1]{inputenc}
\usepackage[english]{babel}
\usepackage{amsmath}
\usepackage{braket}
\usepackage{overpic}
\usepackage{amsfonts}
\usepackage[colorlinks=true,linkcolor=blue, citecolor=blue,allcolors=blue]{hyperref}%
\usepackage{amssymb}
\usepackage{bbold}
\usepackage{makeidx}
\usepackage{natbib}
\usepackage{graphicx}
\usepackage{xcolor}
\usepackage{csquotes}
\usepackage{lmodern}
\usepackage{bbm}

\DeclareMathOperator{\Tr}{Tr}

\begin{document}
\title{Noise-assisted and monitoring-enhanced quantum bath tagging}

\author{Donato Farina}
\affiliation{ICFO - Institut de Ciencies Fotoniques, Mediterranean Technology Park, 08860 Castelldefels (Barcelona), Spain}

\author{Vasco Cavina}
\affiliation{Complex Systems and Statistical Mechanics, Physics and Materials Science,
University of Luxembourg, L-1511 Luxembourg, Luxembourg}

\author{Marco G. Genoni}
\affiliation{Quantum Technology Lab, Dipartimento di Fisica Aldo Pontremoli, Universit\'{a} degli Studi di Milano, I-20133 Milano, Italy}

\author{Vittorio Giovannetti}
\affiliation{NEST, Scuola Normale Superiore and Istituto Nanoscienze-CNR, I-56127 Pisa, Italy}

%

\date{\today}
\begin{abstract}
We analyze the capability of discriminating the statistical nature of a thermal bath by exploiting the interaction with an additional environment. We first shows that, at difference with the standard scenario where the additional environment is not present, the  modified evolution induced by the mere
presence of the extra bath allows to improve the discrimination task. We then also consider the possibility of continuously monitoring the additional environment and we discuss in detail how to obtain improved performances in the discrimination by considering different kinds of interaction, i.e. different jump operators, and different monitoring strategies corresponding to continuous homodyne and photo-detection. Our strategy can be in principle implemented in a circuit QED setup and paves the way to further developments of quantum probing via continuous monitoring.
\end{abstract}

\maketitle

\section{Introduction}
In quantum metrology and quantum sensing 
\cite{giovannetti2006qmetrology,giovannetti2011advances,paris2009qestimation,Demkowicz-Dobrzanski2015a,DegenRMP,PirandolaReview} a quantum probe is
any physical system that allows for
the recovery an unknown classical parameter that has been ``attached" to its state
via some dedicated dynamical process.
For instance 
quantum probes have been used to estimate  parameters related either to their Hamiltonian  (e.g. a frequency or a coupling constant) or to their unitary evolution 
 (say a dynamical phase accumulated while moving along a certain  trajectory). 
 Quantum probes have also been exploited in order to reconstruct the properties of the surrounding environment~\cite{Mehboudi2019Review}; examples are protocols of quantum thermometry \cite{Stace2010,BrunelliPRA2011,correa2015thermometry,Paris_2015,depasquale2016local,depasquale2017temperature,Campbell2018,KiilerichThermo,CavinaPRA2018,Razavian:2019aa}  or aimed to characterize the spectrum of the environment itself \cite{benedetti2018spectral,Bina2018}. More recently it has been proposed to use a quantum probe to discriminate between thermal baths characterized by different thermal~\cite{candeloro2021} or statistical~\cite{farina2019tagging,gianani2020discrimination} properties. In particular in the latter scenario, a quantum probe $S$ is exploited to determine whether the the thermal bath $E$  obeys to  Bosonic or Fermionic statistics, a task which hereafter will be referred to as Quantum Bath Tagging (QBT).  In such scheme  $S$ is let to weakly interact  with $E$ for some time $t$ and then measured 
using optimal detection procedures identified by solving the associated 
 quantum hypothesis testing problem~\cite{helstrom1969quantum}.
To improve further the discrimination performances reported in~\cite{farina2019tagging,gianani2020discrimination} and to refer more closely to realistic experimental setups, 
we consider here the possibility that while coupled with $E$, the probe $S$ could be made interact also with a second auxiliary bath $A$
which, at variance with what happens with $E$, is assumed to have known statistical and thermodynamical  properties (specifically we shall take $A$ to be a, zero-temperature, multi-mode electromagnetic (e.m.) field). 
The role of such an extra environment is twofold: on one hand, 
 the presence of $A$ is  used as a way to positively interfere with the $S$-$E$ coupling
 in an effort to 
increase the distinguishability  among the quantum trajectories associated with the two hypothesis of the problem; on the second hand, $A$ is  employed to set up an indirect, continuous monitoring of the evolution of $S$, hence allowing us to 
acquire information about $E$ in real time and not just at the end of the interaction interval. Continuous monitoring of quantum systems \cite{wiseman2009quantum,jacobs2006straightforward} has indeed been proven useful in the context of quantum metrology: in particular several works have either discussed the fundamental statistical tools to assess the precision achievable in this framework~\cite{Guta2007a,Tsang2011,Tsang2013a,molmer2013bayesian,GammelmarkQCRB,Guta2016,Genoni2017,Albarelli2017a}, and presenting practical estimation strategies~\cite{Gambetta2001,Geremia2003,Molmer2004,Stockton2004,Tsang2010,Wheatley2010,Yonezawa2012,Six2015,KiilerichHomodyne,Cortez2017,Ralph2017,Atalaya2017,Albarelli2018Quantum,Shankar2019,Rossi2020PRL,fallani2021learning}. The theoretical framework needed to assess hypothesis testing protocols has been put forward first by Tsang \cite{TsangContinuousHT} and then by Kiilerich and Molmer \cite{kiilerich2018hypothesis}. We will exploit these techniques for our specific aim and we will discuss how and when continuous monitoring can be useful for QBT. 
\\

\par
In Sec.~\ref{sec:discriminationA} we introduce the QBT problem, presenting the physical setup and
discussing  how to assess hypothesis testing in continuously monitored quantum systems. In Sec.~\ref{sec:analysis} we show our main results, starting from the noise-assisted case, to the scenario where we also allow continuous measurements on the additional environment. In Sec.~\ref{sec-experimental-realizations} we discuss possible implementations of our protocol and  draw our conclusions.

\section{The model}
\label{sec:discriminationA}

\begin{figure}
\centering
\includegraphics[width=.8\linewidth]{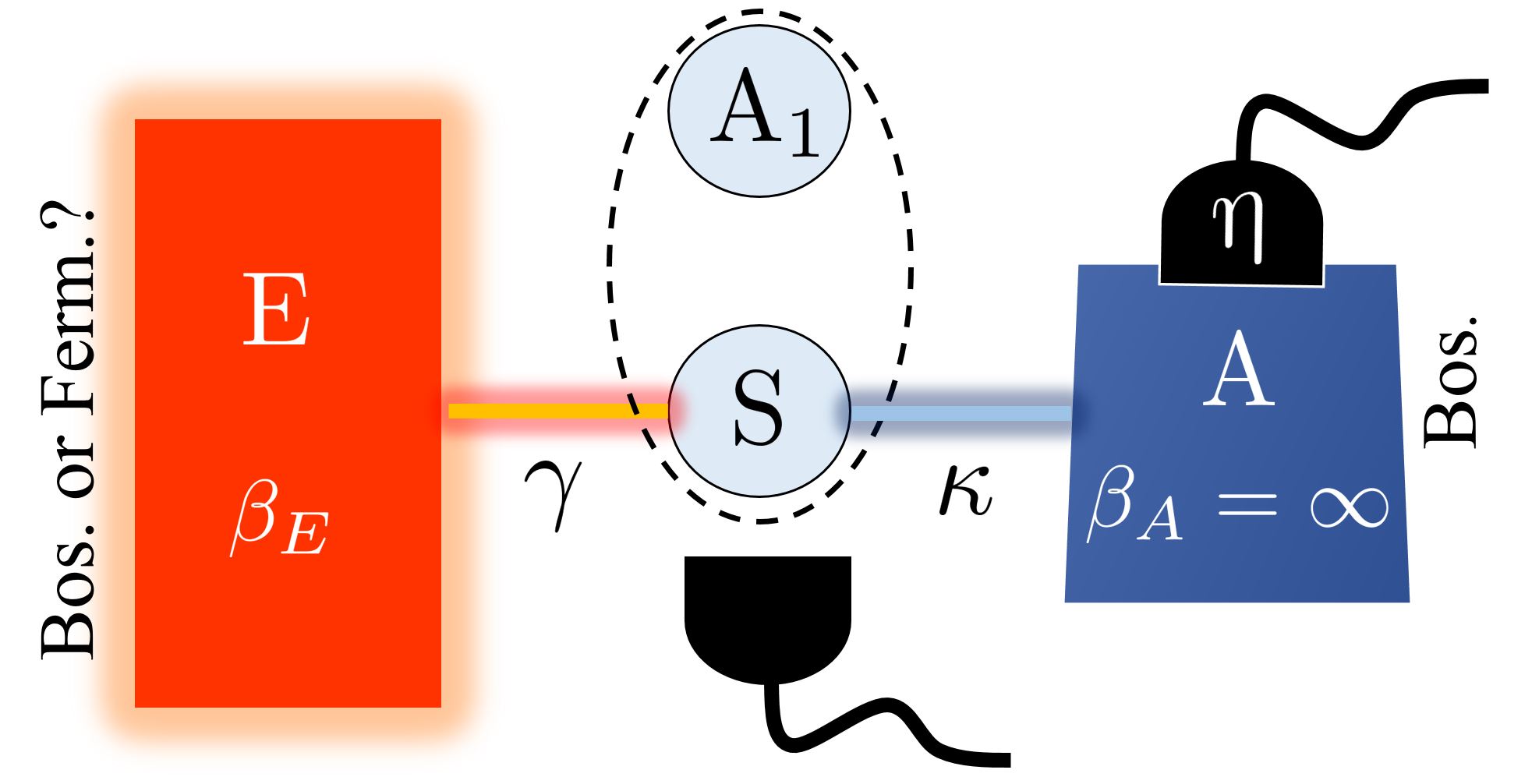}
\caption{Schematic representation  of the QBT setup: 
the statistical nature (Bosonic or Fermionic) of the thermal bath $E$ is determined studying
the modifications it induces on  
 a quantum probe $S$ (a qubit) that has been put in thermal contact with it, while 
  interacting 
 with a zero-temperature Bosonic auxiliary bath $A$ that is  continuously monitored in time
via photo-detection or homodyne measurements. In the picture 
$\gamma$, $\kappa$ are the two decay rates of the unconditional evolution of $S$ for the interaction with $E$ and $A$, respectively,
while $\eta$ is the quantum efficiency of the continuous detection on $A$. 
We also allow for the possibility of initially entangling $S$ with an external qubit memory $A_1$ dynamically decoupled from all the other components of the setup, and performing joint detection on the system $S A_1$.   
\label{schema}}
\end{figure}

The QBT model we study is schematically sketched in Fig~\ref{schema}.
A part from  $E$ (the thermal environment whose statistical nature we wish to determine) and  $S$ (the quantum probe that is put in interaction with $E$), it includes 
two extra elements which were not present in the original QBT scheme discussed in Refs.~\cite{farina2019tagging,gianani2020discrimination}:
namely an auxiliary bath $A$  whose statistical  and thermodynamics properties  are assumed to be known and which is also  attached to $S$,  and an external quantum memory 
 $A_1$ that is dynamically decoupled from all the other components of the setup.
As in Ref.~\cite{farina2019tagging,gianani2020discrimination}, our goal  is to decide whether $E$ is  a Bosonic
bath with assigned inverse temperature $\beta_B$ (hypothesis ${B}$), or Fermionic with assigned temperature $\beta_F$ (hypothesis ${F}$), the initial priors of these two alternatives  being flat. To solve such a task we are
allowed to prepare $S$ (that for simplicity we assume to be a qubit)  in any desired input configuration, possibly correlated with the memory $A_1$, let it evolve for some time $t$ and perform measurements during and/or at the end of the process. 
The possibility of employing  correlated states of $S$ and $A_1$ was not exploited in 
Ref.~\cite{farina2019tagging,gianani2020discrimination} and as we shall see allows for some useful technical improvements. The main difference of our  proposal however is the
presence of the auxiliary bath $A$ which we schematize as a zero-temperature multi-mode e.m. (hence Bosonic) field. Its role is to induce positive interference effects on the $S$-$E$ 
coupling and to permit 
continuous monitoring in time of the system evolution via photo-detection or
or homodyne measurements (a configuration which may physically correspond to the case where we
put $S$ into dispersive QED cavity). 

\subsection{Dynamical evolution} 
In this section we derive the dynamical equations that determine the temporal evolution of
the system. \\

Let us start first by considering the case where the probe interacts with $E$ and $A$ in the absence of 
continuous monitoring of the latter. Following Ref.~\cite{farina2019tagging} we 
model  the $S$-$E$ and $S$-$A$ couplings 
via a 
Gorini-Kossakowski-Sudarshan-Lindblad (GKSL)  master equation ~\cite{GKS1976,lindblad1976generators}, 
 a situation realized under the weak-coupling and Markovian hypotheses \cite{breuer2002theory}. 
Accordingly, 
defining $\mathcal{D}_{[\hat{\theta}]}$ to be the dissipative superoperator
\begin{eqnarray}
\mathcal{D}_{[\hat{\theta}]} \bullet := \hat{\theta} \bullet \hat{\theta}^\dag - \frac{1}{2} \left\{\hat{\theta}^\dag \hat{\theta}, \bullet
\right\} \label{D-operator}
~,
\end{eqnarray}
we write the dynamical evolution of the  joint density matrix ${\hat{\varrho}}_{q}(t)$ of the probe $S$ and the memory $A_1$  as 
\begin{align}
    \frac{d{\hat{\varrho}}_{q}(t)}{dt} = \mathcal{L}_{q}{\hat{\varrho}}_{q}(t) + \kappa \mathcal{D}_{[\hat{c}]} {\hat{\varrho}}_{q}(t)~,
    \label{eq:ME}
\end{align}
where the index $q\in \{ B, F\}$ is used to specify which hypothesis has been selected  for the statistical nature of~$E$.
In this equation $\mathcal{L}_{q}$ is the GKSL dynamical generator which accounts for the free evolution and for the $S-E$ coupling, i.e. \begin{equation}
\label{generator-E}
 \mathcal{L}_{q}\bullet :=-i [\hat{H}_S,\bullet ]
+\gamma [ 1+ s_{q} N_{q}(\beta_{q}) ] 
\mathcal{D}_{[\hat{\sigma}_-]}\bullet
+
\gamma  N_{q}(\beta_{q})  
\mathcal{D}_{[\hat{\sigma}_+]}\bullet,    
\end{equation}
where  $\hat{H}_S:=\omega_0 \hat{\sigma}_+\hat{\sigma}_-$ is the Hamiltonian of the probe,
$\gamma$ is a positive coupling constant that fixes the timescale of the $S$-$E$ interaction, and 
where, having set $s_{q=B}= 1$ and $s_{q=F}=-1$, $N_{q}(\beta):=1/({e^{\beta \omega_0}-s_{q}})$ is the Bose-Einstein/Fermi-Dirac factor: notice that  $\hbar$ has been set equal to $1$ and that no free Hamiltonian has been assumed for $A_1$ which effectively participates to the process only through initial correlations with $S$ that have been  possibly established at the beginning of the dynamical evolution.  
The second term in the l.h.s of Eq.~(\ref{eq:ME}) represents instead the
$S$-$A$ coupling with the operator $\hat{c}$ selected depending on the type of interaction  one has engineered, and with 
$\kappa\geq 0$ being a parameter that gauges its intensity -- in particular setting $\kappa=0$ we recover
the model discussed in Refs.~\cite{farina2019tagging, gianani2020discrimination}. 
 In the following we will consider the two cases $\hat{c} = \hat{\sigma}_-$ and $\hat{c} = \hat{\sigma}_x/2$: the first one corresponds to the a purely dissipative model where $S$ looses energy in favour to $A$ via spontaneous emission, while the second choice can be obtained via dispersive coupling that can be engineered e.g. in circuit-QED systems~\cite{hacohen-gourgy_quantum_2016,chantasri_simultaneous_2018,hacohen-gourgy_continuous_2020}. 
\\

As already mentioned, Eq.~(\ref{eq:ME}) does not include effects associated with a continuous monitoring of $A$. To account for the latter we resort to 
 the stochastic master equations (SME) approach of Refs.~\cite{wiseman2009quantum,jacobs2006straightforward}. In particular we will focus on two kind of measurements, photodetection and homodyne detection with a fixed monitoring efficiency~$\eta$. In the case of photodetection, under  hypothesis $q$, the corresponding SME for the conditional state of $SA_1$  reads
\begin{align}
\label{eq:SMEpd}
    d{\hat{\varrho}}^{c}_{q}(t) &= \mathcal{L}_{q}{\hat{\varrho}}_{q}^{c}(t) \,dt 
    + (1-\eta) \kappa \mathcal{D}_{[\hat{c}]}{\hat{\varrho}}_{q}^c(t)  \\
    &\, - \frac{\eta\kappa}{2} \mathcal{H}_{[\hat{c}^\dag \hat{c}]} {\hat{\varrho}}^{c}_{q}(t) \,dt  \nonumber
    + \left( \frac{\hat{c} {\hat{\varrho}}^{c}_{q}(t) \hat{c}^\dag}{\Tr[\hat{c} {\hat{\varrho}}^{c}_{q}(t) \hat{c}^\dag]} - {\hat{\varrho}}^{c}_{q}(t) \right) dN_t \,,
\end{align}
where $dN_t\in\{ 0,1\}$, corresponds physically to the number of photons detected at each time $t$ and mathematically to a Poisson increment defined by its probability of taking value equal to one, $p(dN_t=1) =\eta \kappa \Tr[{\hat{\varrho}}^{c}_{q}(t) \hat{c}^\dag \hat{c}]\,dt$, and where we have introduced the superoperator
\begin{align}
\mathcal{H}_{[\hat{\theta}]} \bullet := \hat{\theta}
 \bullet + \bullet \hat{\theta}^\dag - \Tr[(\hat{\theta}^\dag + \hat{\theta}) \bullet] \bullet \,.
\end{align}
Similarly, in the case of homodyne detection one obtains the SME
\begin{align}
    d{\hat{\varrho}}^{c}_{q}(t) = \mathcal{L}_{q}{\hat{\varrho}}_{q}^{c}(t) \,dt 
  +  \kappa \mathcal{D}_{[\hat{c}]} {\hat{\varrho}}^{c}_{q} (t) \,dt 
+ \sqrt{\eta\kappa} \mathcal{H}_{[\hat{c}]}{\hat{\varrho}}^{c}_{q}(t) \,dW_t \,,
  \label{eq:SMEhd}
\end{align}
where the state is conditioned on the continuous output photocurrent
\begin{align}
dy_t := \sqrt{\eta\kappa} \Tr[{\hat{\varrho}}^{c}_{q}(t) (\hat{c} + \hat{c}^\dag)] \,dt + dW_t \,, 
\label{eq:photocurrent}
\end{align}
and where $dW_t$, denoting the difference between the measurement output $dy_t$ and the expected results, mathematically corresponds to a Wiener increment s.t. the relation $dW_t^2 =dt$ holds deterministically. We remark that by choosing as jump operators the ones defined before, i.e. either $\hat{c} = \hat{\sigma}_-$ or $\hat{c} = \hat{\sigma}_x/2$, one obtains photocurrents \eqref{eq:photocurrent} with the same form, yielding information on the average value of the operator~$\hat{\sigma}_x$. However the two operators will induce different dynamics, described by the corresponding SMEs \eqref{eq:SMEhd}. 

For both photodetection and homodyne detection strategies, 
the associated  SME  (\ref{eq:SMEpd}) and (\ref{eq:SMEhd}) can be numerically integrated
following the  method based on Kraus operators suggested in \cite{rouchon2014models,rouchon2015efficient} that we review in brief in Appendix \ref{a:numerical}. This results in a collection of 
quantum trajectories for the conditional density matrix $\hat{\varrho}_{q}^{c}(t)$ each identified by a string of records
\begin{equation} \label{eq:string1}
D_t:=(x_{t_0+dt},x_{t_0+2 dt}, \dots, x_{t-dt}, x_{t})~,
\end{equation}
where we have assumed to perform measurements every infinitesimal time-interval $dt$ starting from the initial time $t_0$ (which we set to zero hereafter) and stopping at time $t$, and where for the  photodetection and homodyne detection scenario the $x_t$'s  
 correspond either to recorded values of $dN_t$ or $dy_t$ respectively. 
 It is worth pointing out that, in principle,
  by averaging 
 over all such solutions, i.e. by averaging $\hat{\varrho}_{q}^{c}(t)$ over all the obtained measurement results (\ref{eq:string1}) up to a time $t$ or equivalently by fixing the monitoring efficiency $\eta=0$, one obtains  an unconditional  state solution that coincides with  the standard master equation (\ref{eq:ME}) of the problem, i.e. 
 \begin{eqnarray} \label{AVEAWAY} 
\mathbbm{E}[\hat{\varrho}_q^c(t)]=\left. \hat{\varrho}_q^c(t)\right|_{\eta=0} = \hat{\varrho}_q(t) \;. 
 \end{eqnarray}

\subsection{Quantum hypothesis testing in continuously monitored quantum systems}
\label{sec:hyptesting}

In this section we  review the methods that allow us to characterize how efficiently one can solve the QBT problem we are facing. 
\\

To begin with, consider first  the simple case where the data from the continuous monitoring in time are neglected, e.g. by averaging them away or setting $\eta=0$, a regime in which thanks to~(\ref{AVEAWAY}) the evolution of the system is provided by 
the master equation~(\ref{eq:ME}).
Having hence selected an input state ${\hat{\varrho}}(0)$ for the $SA_1$ system and a total evolution time $t$, what we have to do is to determine whether at the end of the process the state of $SA_1$ is better described by the density matrix ${\hat{\varrho}}_{B}(t)$ or by the density matrix  ${\hat{\varrho}}_{F}(t)$ obtained by solving
 Eq.~(\ref{eq:ME}) under the two alternative QBT hypotheses. This problem can be easily framed as a special instance of quantum hypothesis testing~\cite{helstrom1969quantum}: accordingly we can  bound  the error probability associated with the selected strategy  through the  Helstrom inequality
\begin{eqnarray}\label{HEPBOUND} 
p_{err}(t;{\hat{\varrho}}(0))  &\geq& {\rm HEP}(t;{\hat{\varrho}}(0))\;, 
\\
\label{Helstrom} 
{\rm HEP}(t;{\hat{\varrho}}(0)) &:=& \frac{1}{2} \left(1- \frac{ \|  {\hat{\varrho}}_{B}(t)- {\hat{\varrho}}_{F}(t)\|_1}{2}\right) ,  
\end{eqnarray} 
where $\lVert \bullet \lVert_1$ denotes the trace norm, and where we used the fact that
the prior probability associated with the events $B$ and $F$ is flat.
The threshold  value ${\rm HEP}(t;{\hat{\varrho}}(0))$, conventionally called the Helstrom error probabilitiy,  
can always be attained via a projective measurement on $SA_1$ 
that at time $t$ distinguishes the positive and negative eigenstates of the operator ${\hat{\varrho}}_{B}(t)- {\hat{\varrho}}_{F}(t)$. Accordingly in Refs.~\cite{farina2019tagging,gianani2020discrimination} ${\rm HEP}(t;{\hat{\varrho}}(0))$ was
 used as a bona-fide quality factor for the QBT efficiency one can achieve with the selected choice of $t$ and 
 ${\hat{\varrho}}(0)$. Notice however that in such works  ${\hat{\varrho}}_{B}(t)$ and ${\hat{\varrho}}_{F}(t)$
 referred to the local states of~$S$ (i.e. the presence of the external quantum memory $A_1$ was not allowed) and,  most importantly,   the auxiliary bath $A$ was not included in the picture (a condition which in our modelization  corresponds to set $\kappa=0$ in Eq.~(\ref{eq:ME})).
As we shall see in the next section, even without resorting to continuous monitoring in time,  lifting these two constraints already allows one for some non trivial improvements on the minimum error probability value. 
\\

Let's now address the QBT problem and continuous monitoring assumptions. As described in \cite{kiilerich2018hypothesis}, in this case the hypothesis testing can follow two different approaches: in order discriminate between the two hypotheses, one may exploit the continuous experimental data $D_t$ only, or one can also implement a final direct measurement on $S$ and $A_1$ on the corresponding conditional states. We now start to asses the first scenario. In this case one can resort to a Bayesian analysis, by first observing that each trajectory $D_t$ is characterized by a probability $P(D_t | q)$, when conditioning on the initial assumption that the bath is defined by a statistics associated with the QBT hypothesis $q$. Hence, introducing a likelihood 
$L(D_t | q ) =P(D_t | q )/p_0(D_t)$  with $p_0(D_t)$ denoting a positive function of $D_t$ only~\cite{kiilerich2018hypothesis}, and by resorting to Bayes theorem, it is possible to compute the {\em a-posteriori} probability as 
\begin{align}
P(q | D_t ) = \frac{P(D_t | q)}{\sum_{q^\prime} P(D_t | q^\prime)} =  \frac{L(D_t | q)}{\sum_{q^\prime} L(D_t | q^\prime)} \,, \label{eq:bayes}
\end{align}
which we present here exploiting the fact that the  flat prior distribution on $q$ is flat
(the specific definition of  $L(D_t | q )$  and the method to efficiently compute it is discussed in details in Appendix \ref{a:numerical}).
Observe next that as~\eqref{eq:bayes} is normalized for each values of the QBT hypothesis index $q$ we have two possibilities, namely
$P(q|D_t) \geq 1/2$, in such a case the bath is most likely to be of $q$ nature, and $P(q|D_t) < 1/2$ in which the opposite hypothesis is more plausible.
However, the inherent stochasticity of the measurement outcomes can result in $P(B|D_t) \geq 1/2$ ($P(F|D_t) \geq 1/2$) even if the statistics of the bath was Fermionic (resp. Bosonic), i.e. there are measurement records that may lead to a wrong inference process. The goal is thus to quantify the probability of occurrence of such wrong tagging events.
In the spirit of a thought experiment, we consider a sample of $N_{traj}$ trajectories $D_t$, supposing that half of them are generated by indirectly probing a Bosonic environment ($ D^{B}_t $), while the rest are Fermionic ($ D^{F}_t $).
A wrong tagging event is triggered every time we have a trajectory $D^{q}_t$ such that $P(q| D^{q}_t) < 1/2$. Counting the number $N_{wrong}$  of such trajectories leads to a first way to quantify the error probability as the following ratio
\begin{equation}
p_{err}^{(cont)}(t;{\hat{\varrho}}(0)) :=
\frac{N_{wrong}}{N_{traj}}~,
\label{err-cont}
\end{equation}
where the notation stresses the implicit dependence upon the specific choice of the input state ${\hat{\varrho}}(0)$ of $SA_1$ and on the total evolution time $t$.

As mentioned before, a discrimination capability higher than (\ref{err-cont}) 
can in principle be achieved by improving our continuous monitoring scheme with the addition of a Helstrom projective measurement 
on $S$ and $A_1$ at the final time $t$. In this case the ultimate bound for the error probability is given by the Helstrom bound \eqref{Helstrom}, for the two quantum states $\hat{\varrho}_{B,F}^{c}(t)$, solutions of the SMEs \eqref{eq:SMEpd} or \eqref{eq:SMEhd} for the dataset $D_t$, and obtained numerically via Eq.~\eqref{eq:rouchonralph}, with prior probabilities $P(q|D_t)$. In formula we obtain the following non linear functional of the detector records
%
\begin{equation}
\label{hel-traj}
{\rm HEP}^{c}(t;\hat{\varrho}(0))
:=\frac{1-
\left\lVert
P(F|D_t)\hat{\varrho}_F^{c}(t)
-
P(B|D_t)\hat{\varrho}_b^{c}(t)
\right\lVert_1}{2} ~.
\end{equation}
%
We remark that, apart influencing the dynamics of the density matrices $\hat{\varrho}_b^{c}(t)$ and $\hat{\varrho}_F^{c}(t)$, the knowledge coming from continuous monitoring updates the two priors probabilities \cite{kiilerich2018hypothesis}, and in general indentifies the optimal Helstrom projective measurement.
An average over all the $N_{traj}$ trajectories of our sample returns the following figure of merit
\begin{eqnarray}
p_{err}^{( cont +  proj)}(t;\hat{\varrho}(0)):=\mathbbm{E}\left[{\rm HEP}^{c}(t;\hat{\varrho}(0))
\right]~,
\label{err-cont+proj}
\end{eqnarray}
that thus takes into account the average information gained from both the continuous monitoring and from a final Helstrom projective measurement for each trajectory.
\section{Analysis/Results}
\label{sec:analysis}
In this section we will present our main results. We will start by discussing the standard QBT scenario presented in \cite{farina2019tagging}, but allowing the system to be entangled with a qubit ancilla. We will then
study how the presence of the auxiliary bath $A$ can be useful in  solving the  quantum hypothesis testing
problem even in the absence of the continuous monitoring (an effected which we can dub noise-assisted scenario).
Finally we address 
 the continuous-monitoring scenario showing how the indirect information obtainable from $A$ could help in the discrimination strategy. 
\subsection{Advantage from initial entanglement}
\label{sec:extended-scenario}
As already discussed in the previous sections in the original QBT schemes of Refs.~\cite{farina2019tagging, gianani2020discrimination} the probing system $S$ was not correlated with  external memory elements. 
We argue here that adding $A_1$ into the picture already introduces some major advantages, even in the absence of the extra auxiliary bath $A$ and of its continuous monitoring in time. 
To see this explicit let us introduce ${\cal E}_{q,t}$ the Linear, Completely Positive, Trace Preserving (LCPT) channel \cite{watrousBook,HolevoBook}
that allows one to express the solution of Eq.~(\ref{eq:ME}) as $\hat{\varrho}_{q}(t) = {\cal E}_{q,t}\hat{\varrho}(0)$.
 Observe hence that, for fixed $t$, the minimal value that the HEP function of Eq.~(\ref{HEPBOUND}) can attain  can be expressed as  \begin{equation}\label{HEPBOUNDminimal} 
 {\rm HEP}(t;{\hat{\varrho}}(0))\geq  {\rm HEP}_{\diamond}(t):=
 \frac{1}{2} \left(1- \frac{ \|  {\cal E}_{B,t}-{\cal E}_{F,t}  \|_\diamond}{2}\right) ,  
\end{equation} 
with \begin{equation}
\|  {\cal E}_{B,t}-{\cal E}_{F,t}  \|_\diamond:= \max_{\hat{\varrho}(0) \in \mathfrak{S}_{SA_1}} 
\|  {\cal E}_{B,t}\hat{\varrho}(0)  -  {\cal E}_{F,t} \hat{\varrho}(0) \|_1\;, 
  \label{dNORM} 
\end{equation} 
being the 
 the  diamond norm  distance~\cite{Kitaev_1997,KitaevBook} obtained by  maximizing  over the set 
 of $\mathfrak{S}_{SA_1}$ of the input  joint  density matrices of
 $S$ and $A_1$.
 The term ${\rm HEP}_{\diamond}(t)$ of (\ref{HEPBOUNDminimal}) should be compared 
 with the quantity
 \begin{eqnarray}
 {\rm HEP}_{1}(t):= \frac{1}{2} 
 \left(1- \frac{ \|  {\cal E}_{B,t}-{\cal E}_{F,t}  \|_1}{2}\right)\;, 
 \end{eqnarray} 
  with
 \begin{eqnarray} \|  {\cal E}_{B,t}-{\cal E}_{F,t}  \|_1:= \max_{\hat{\varrho}(0) \in \mathfrak{S}_S} 
\|  {\cal E}_{B,t}\hat{\varrho}(0)  -  {\cal E}_{F,t} \hat{\varrho}(0) \|_1,\end{eqnarray} 
  which represents instead the optimal QBT error probability  
 one can get 
  by restricting the analysis to
 only local density matrices of $S$ as assumed in Refs.~\cite{farina2019tagging, gianani2020discrimination}. The fact that using $A_1$ can provide better QBT performances then follows simply by  the natural ordering between the diamond norm distance and the corresponding trace norm distance~\cite{watrousBook},
which implies $\|  {\cal E}_{B,t}-{\cal E}_{F,t}  \|_\diamond \geq \|  {\cal E}_{B,t}-{\cal E}_{F,t}  \|_1$, and hence 
\begin{eqnarray} {\rm HEP}_{1}(t)\geq {\rm HEP}_{\diamond}(t)\;. 
\label{ADV}\end{eqnarray}  
A quantitive evaluation of  the advantages implied by Eq.~(\ref{ADV}) can be attained 
by focusing on the scenario where $A$ is disconnected (i.e. $\kappa=0$) and  the temperature of $E$ in two QBT hypothesis is the same, i.e. $\beta_B=\beta_F= \beta$, and sufficiently large, i.e. $\beta\rightarrow 0$. 
 Under these conditions,  the rescaled rate constants corresponding to the Bosonic hypothesis, i.e. $q=B$, diverge while those for the Fermionic hypothesis, i.e. $q=F$, remain finite. 
 As a consequence the Bosonic channel ${\cal E}_{B,t}$ will imply immediate thermalization of $S$, 
in the sense that it leads to thermalization of the probe  system on time scales $\tau$  where the Fermionic channel ${\cal E}_{F,t}$ has not significantly affected the dynamics yet, i.e.
formally  
\begin{eqnarray}
{\cal E}_{B,\tau} \bullet\simeq \hat{\varrho}_{\beta} \otimes \mbox{Tr}_S[\bullet]  \;, \qquad 
{\cal E}_{F,\tau}\bullet\simeq {\cal I}\bullet \;, 
\end{eqnarray} 
with 
\begin{eqnarray} \label{GIBBS} 
\hat{\varrho}_{\beta}:=\exp(-\beta \hat{H}_S)/{\rm tr}[\exp(-\beta \hat{H}_S)]\;,\end{eqnarray}  the Gibbs thermal state of the probe, $\mbox{Tr}_S[\bullet]$ represents the partial trace with respect to $S$, and 
${\cal I}$ is the identity superoperator. 
Choosing hence the initial state $\hat{\varrho}(0)$ of $S$ and $A_1$ to be the maximally entangled state 
\begin{eqnarray} \label{MAXent} 
\ket{\Phi^+}:=(\ket{11}_{SA_1}+\ket{00}_{SA_1})/\sqrt{2}\;,\end{eqnarray}  from (\ref{Helstrom}) we get 
\begin{eqnarray}
\|  {\cal E}_{B,\tau}\hat{\varrho}(0)  -  {\cal E}_{F,\tau} \hat{\varrho}(0) \|_1&\simeq& 
 \left\|  {\hat{\varrho}}(0) - 
\hat{\varrho}_{\beta} \label{TRACEdist} 
\otimes \frac{\mathbb{1}}{2}  \right\|_1 \\
&\simeq& \nonumber
\left\|  \ket{\Phi^+}\bra{\Phi^{+}}  - 
\frac{\mathbb{1}}{2} 
\otimes \frac{\mathbb{1}}{2}  \right\|_1 =\frac{3}{2}\;, \end{eqnarray}
where in the second line we invoke the $\beta\rightarrow 0$ limit to approximate 
$\hat{\varrho}_{\beta}\simeq \mathbb{1}/{2}$. Replacing this into 
 Eq.~(\ref{Helstrom}) gives finally 
\begin{eqnarray}
{\rm HEP}(\tau;\ket{\Phi^+})\simeq1/8\;, \label{BEST} 
\end{eqnarray}  which should  be compared with the values one would get by discarding $A_1$ from the problem, i.e. forcing $\hat{\varrho}(0)$ to be a local density matrix of just $S$. Under this circumstance, Eq.~(\ref{TRACEdist}) gets replaced by 
\begin{equation}
\|  {\cal E}_{B,\tau}\hat{\varrho}(0)  -  {\cal E}_{F,\tau} \hat{\varrho}(0) \|_1\simeq
 \left\|  {\hat{\varrho}}(0) - 
\hat{\varrho}_{\beta} \label{TRACEdistnew} 
  \right\|_1
\simeq
\| {\hat{\varrho}}(0)  - 
{\mathbb{1}}/{2} 
  \|_1 \leq 1\;, \end{equation}
with  the last inequality being reachable by taking $\hat{\varrho}(0)$ pure, e.g. the vector $|1\rangle$.
Accordingly we can write
\begin{eqnarray}
 {\rm HEP}_{1}(\tau)= {\rm HEP}(\tau; |1\rangle) \simeq 
  {1}/{4} \;, 
 \end{eqnarray}
which is twice the minimum value~(\ref{BEST})  attained by using as input for $S$ and $A_1$ the maximally entangled state~(\ref{MAXent}). 

\subsection{Noise-assisted QBT}

In this section we show that 
the presence of the additional environment $A$ can enhance the QBT procedure even when $A$ is not continuously monitored, i.e. for $\kappa>0$ but $\eta=0$ so that  the equation of motion of the system is governed by Eq.~(\ref{eq:ME}). Since in this case $A$ introduces only extra  dissipative effects in the dynamics of $S$ we dub such enahacement noise-assisted QBT. As we shall see the origin of this phenomenon can be traced back to the fact that 
adding $A$ into the picture (i.e. passing from $\kappa=0$ to $\kappa\neq 0$  in Eq.~(\ref{eq:ME})) 
modifies the dynamical process which is responsible for the encoding of the statistical nature of $E$ on the $SA_1$ system. While in a generic metrology setting there is no guaranty that such interference will have positive effects, the theory does not prevent that for some special task this could happen: 
the QBT problem we present here is one  (indeed to our knowledge probably the first)
of the such special examples. 

\begin{figure}
\centering
\begin{overpic}[width=.8\linewidth]{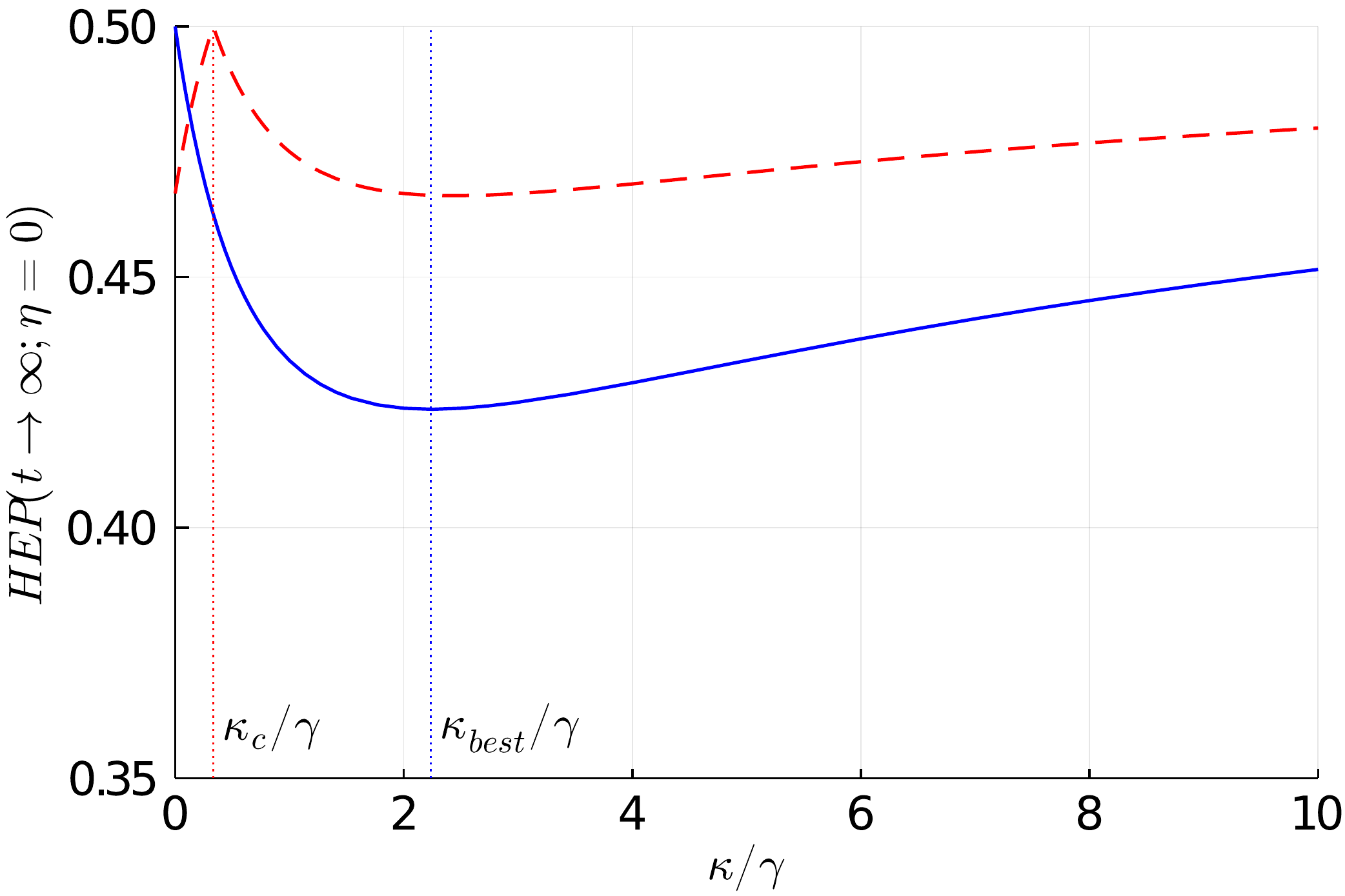}
\put(90,65){}
\end{overpic}
\caption{\label{fig:kappa-load}
Plots of the noise-assisted Helstrom bound ${\rm HEP}(t\rightarrow \infty)$ reported in Eq.~\eqref{hel-0eta}, as function of  $\kappa/\gamma$.
Solid blue line: isothermal QBT scenario  $\beta_B=\beta_F=\beta$ with $N_B(\beta)=2$. 
Dashed red line: example of an asymmetric temperature QBT scenario ($N_B(\beta_B)=1$, $N_B(\beta_F)=2$). Here we notice that ${\rm HEP}(t\rightarrow \infty)$ is smaller than $1/2$  for $\kappa=0$, reaching instead the zero discrimination threshold at an intermediate critical value determined by Eq.~(\ref{critical}).}
\end{figure}

To enlighten the possibility of exploiting the mere presence of $A$ to boost the QBT performances it is useful to consider  the scenario where the two QBT hypothesis 
are characterized  by the same temperature (i.e. $\beta_B = \beta_F=\beta$): under this condition
 for $t$ sufficiently large,  the contact with $E$  alone  ($\kappa =0$) will lead $S$ to the Gibbs state~(\ref{GIBBS}), regardless to the nature of the bath hence making the QBT discrimination impossible~\cite{farina2019tagging}. Yet there is a chance that 
 by taking $\kappa\neq 0$, the simultaneous interactions of $S$ with $E$ and 
$A$ will interfere leading to departures from such dead-end   behaviour paving the way for improvements of the discrimination efficiency even for large $t$ (it is also clear however that one also expect that  in order to be
beneficial, such deviations should not be too strong so that the $S$-$E$ coupling gets completely dominated by the $S$-$A$ interaction). 
To see this explicitly let us study the values that the HEP figure of merit 
${\rm HEP}(t;{\hat{\varrho}}(0))$ of Eq.~(\ref{Helstrom}) attains in the asymptotic regime of $t\rightarrow \infty$ as a function of $\kappa$, 
considering the scenario where  the $S$-$A$ interaction is mediated by the operator $\hat{c} = \hat{\sigma}_-$ (dissipative coupling). In this case, irrespectively from the choice of the initial state ${\hat{\varrho}}(0)$ of $S$ and $A_1$, we obtain the following steady state HEP value
\begin{equation}
{\rm HEP}(t\rightarrow \infty)=\frac{1}{2}-\frac{1}{2 \omega_0 \kappa}\Big|
\dot{Q}^{(E_B \Rightarrow A )} - \dot{Q}^{(E_F \Rightarrow A )}
\Big|~, 
\label{hel-0eta}
\end{equation}
where $\dot{Q}^{(E_q\Rightarrow A)}$ denotes the heat flows from $E$ to $A$ associated with the QBT hypotheses $q=B,F$, i.e. the quantities 
\begin{eqnarray} \label{flow}
&&{ \dot{Q}^{(E_B\Rightarrow A)}=}
\omega_0 \kappa\; 
\frac{N_B(\beta_B)}{1+2 N_B(\beta_B)
+\kappa/\gamma}
~,
\\
&&{ \dot{Q}^{(E_F\Rightarrow A)}=}
\omega_0
\kappa\; 
\frac{N_B(\beta_F)
}{1+2 N_B(\beta_F)
 +[1+2 N_B(\beta_F)
] \kappa/\gamma}
~,\nonumber
\end{eqnarray}
which we report here for arbitrary choices of $\beta_B$ and $\beta_F$.
The result \eqref{hel-0eta} holds also for the case $\hat{c}=\hat{\sigma}_x/2$, up to a numerical factor and different expressions for the heat flows (see Appendix \ref{app:steady} for the derivation of all the results).
As anticipated we notice that for $\beta_B=\beta_F=\beta$ and $\kappa=0$, one gets 
 ${\rm HEP}(t\rightarrow\infty) = 1/2$  signalling the impossibility of solving the QBT problem \cite{farina2019tagging}.
We observe also that  for $\kappa\gg \gamma$ one has 
${\rm HEP}(t\rightarrow\infty) = 1/2$ signalling that the large disturbance originated by $A$ nullifies the sensitivity to the  statistics of the bath $E$. 
Most interestingly however when $\kappa$ is finite we get a clear advantage with respect to the $\kappa=0$ case, see Fig.~\ref{fig:kappa-load}. The physical interpretation of such noise-assisted QBT improvement is that since $A$ is a zero-temperature bath there is 
finite average heat flowing from the hot bath $E$ with ${q}=B,F$ that can be monitored by the probe:
 the non-zero discrimination capability hence follows due to the fact that a Fermionic  $E$ implies a slower heat transfer from $E$ to $S$ than a Bosonic $E$. 
 Figure~\ref{fig:kappa-load} makes also evident that
 there exists in particular an optimal coupling constant $\kappa$ minimizing \eqref{hel-0eta}, that for $\hat{c}=\hat{\sigma}_-$ can be analytically evaluated as 
\begin{equation}
\kappa_{best}/\gamma=\sqrt{2 N_B(\beta)+1}~, \label{Kbest}
\end{equation}
(when the $S$-$A$ is mediated by the operator  $\hat{c}=\hat{\sigma}_x/2$ the optimal value is twice as above -- see Appendix \ref{app:steady}).

A similar analysis can also be conducted for 
the  case of asymmetric temperatures ($\beta_B\neq\beta_F$): here however the model naturally allows also for discrimination at steady state  also in the case $\kappa=0$, as already studied in \cite{gianani2020discrimination}. 
Accordingly, while in some regimes one can still get improvements by working with $\kappa\neq 0$  the study become slightly more involved and possibly less interesting. Instead we would like to report the fact that 
 in this unequal temperature scenario there can be critical $\kappa$ values
\begin{eqnarray}
\kappa_c/\gamma=\frac{N_B(\beta_F)-N_B(\beta_B)}{N_B(\beta_B)[1+2N_B(\beta_F)]-N_B(\beta_F)} \label{critical}
\end{eqnarray}
where QBT discrimination is made impossible (i.e. $H(t \rightarrow \infty)=1/2$) by the presence of $A$, see the dashed line in Fig.~\ref{fig:kappa-load}. For the $\hat{c} = \hat{\sigma}_-$
coupling this is actually happening if and only if
either we have 
\begin{eqnarray} 
&&\frac{1}{2}<N_B(\beta_B)<N_B(\beta_F) \;, 
\end{eqnarray}
or 
\begin{eqnarray} 
&&N_B(\beta_B)<N_B(\beta_F)<\frac{N_B(\beta_B)}{1-2N_B(\beta_B)}~.
\end{eqnarray} 
This last property marks a difference with the noise assisted QBT with $\hat{c}= \hat{\sigma}_x/2$ where  a non zero discrimination capability at steady state for $\kappa$ finite may occur but the critical points appear only for $N_B(\beta_B) \geq N_B(\beta_F)$ (see Appendix \ref{app:steady}).
In summary, the additive noise implied by an engineered additional environment on the one hand can open the discrimination window for two baths at the same temperature, on the other hand can prevent discrimination of two baths at different temperatures when choosing \enquote{unlucky} values of the loss coefficient.

\subsection{Monitoring-enhanced QBT}

\begin{figure}
\vspace{1cm}
\centering
\begin{overpic}[width=.8\linewidth]{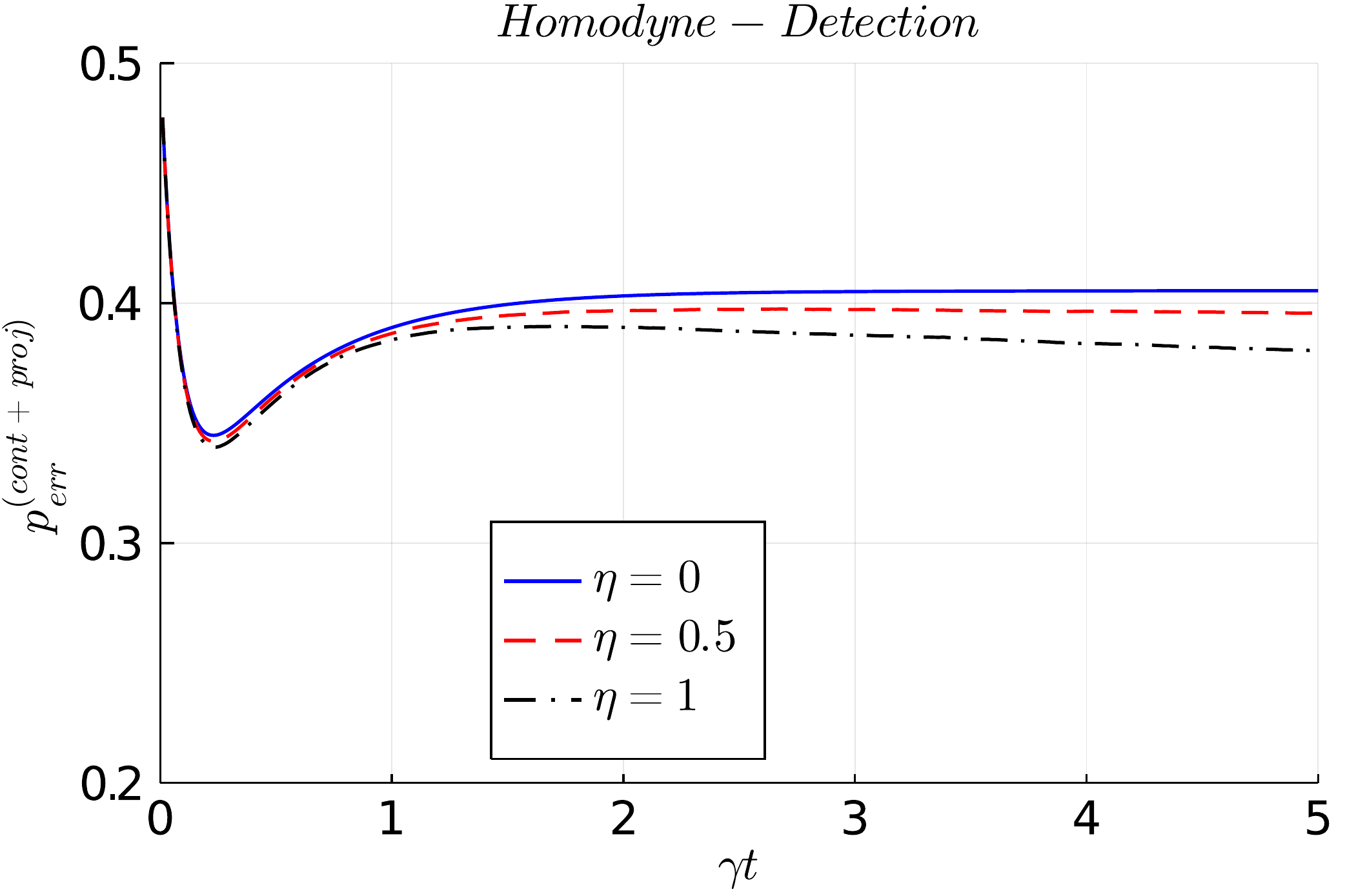}
\put(50,55){\color{black} input $\ket{0}$}
\put(90,65){(a)}
\end{overpic}
\\
\vspace{.5cm}
\begin{overpic}[width=.8\linewidth]{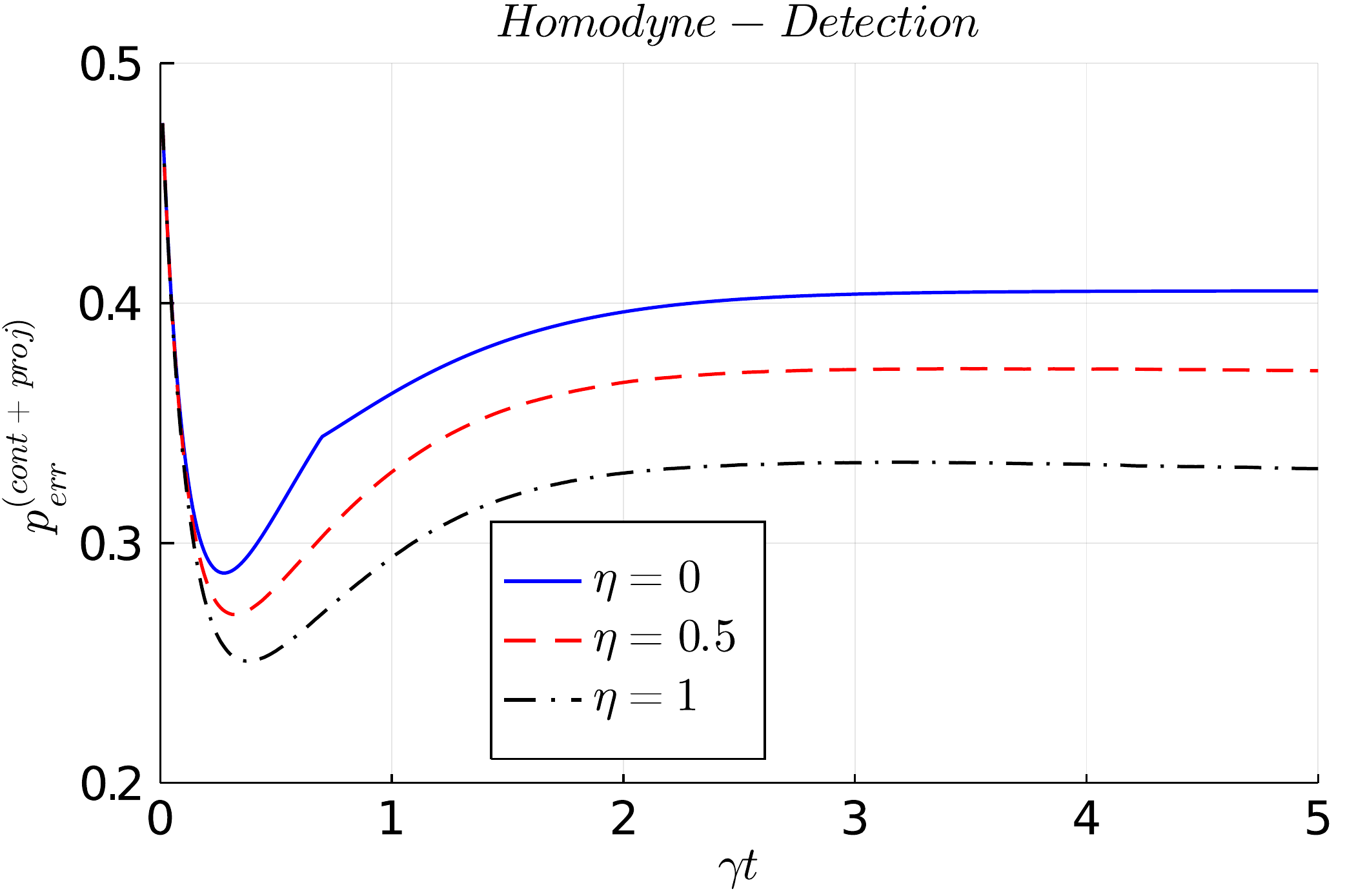}
\put(50,55){\color{black} input $\ket{\Phi^+}$}
\put(90,65){(b)}
\end{overpic}
\caption{\label{fig:vari-eta}
Plot of $p_{err}^{(cont+proj)}$, Eq.~\eqref{err-cont+proj}, for the different values of the efficiency $\eta$ reported in the legend.
In panel
(a) we choose the input state of $S$ to be the ground state of its local Hamiltonian,
while in 
panel (b) 
we consider as input  of $S$ and $A_1$ the maximally entangled state 
$\ket{\Phi^+}$.
All data in the figure are obtained 
in the isothermal QBT scenario  $\beta_B=\beta_F=\beta$,
for 
$\beta \omega_0=1/5.5$; $\kappa/\gamma=1$ and by considering the jump operator $\hat{c}=\hat{\sigma}_-$.}
\end{figure}

We now discuss the performance in the QBT protocol when the additional environment can be continuously monitored, by considering the two scenarios corresponding to either flourescence or dispersive monitoring corresponding respectively to the jump operators $\hat{c} =\hat{\sigma}_-$ or $\hat{c}=\hat{\sigma}_x/2$. We remind that under these circumstances 
the system dynamics is described by the SME~(\ref{eq:SMEpd}) or
(\ref{eq:SMEhd}) depending on the type of measurements we have selected, and that 
 the attainable mean error probability 
can be evaluated either in terms of the functional $p_{err}^{(cont)}(t;{\hat{\varrho}}(0))$ of Eq.~(\ref{err-cont}), or in terms of its improved version $p_{err}^{( cont +  proj)}(t;\hat{\varrho}(0))$ of Eq.~(\ref{err-cont+proj}), depending on whether or not we allow for a final Helstrom measurement on $S$ and $A_1$.
In an effort to simplify the study in what follows we shall fix as input state for $SA_1$ the maximally entangled state~(\ref{MAXent}) -- the only exception being for the data reported in panel (a) of Fig.~\ref{fig:vari-eta} where we assume $S_1$ to be  uncorrelated with $A_1$. While in principle for given $t$ this is possibly not the optimal choice in terms of the diamond norm requirement, the choice is an educated guess as its evolved counterpart is nothing but  the Choi-Jamiolkowski state~\cite{watrousBook,HolevoBook} of the associated dynamical  map that is known to provide a faithful representation of the latter. 

\subsubsection{Purely dissipative  $S$-$A$ coupling regime }
Here we focus on the case where $S$ and $A$ interact through the 
  jump operator $\hat{c} = \hat{\sigma}_-$.
 
The usefulness of exploiting the knowledge deriving from the continuous monitoring of $A$ 
are well enlightened in Fig.~\ref{fig:vari-eta} where for brevity we only focus on homodyne-detection:
in this figure the quantity 
  $p_{err}^{(cont + proj)}(t)$ is plotted as function of $t$, for different choices of the quantum efficiency $\eta$. As intuitively expected increasing  $\eta$ leads to better discrimination performance: in particular the worst case scenario is  obtained for $\eta=0$ (blue curves in the plot, corresponding to the noise-assisted strategy where  we do not monitor~$A$), while the best case is associated with $\eta=1$ (black dash-dotted curve, corresponding to perfect detection efficiency). 

\begin{figure*}
\vspace{1cm}
\centering
\begin{overpic}[width=.3\linewidth]{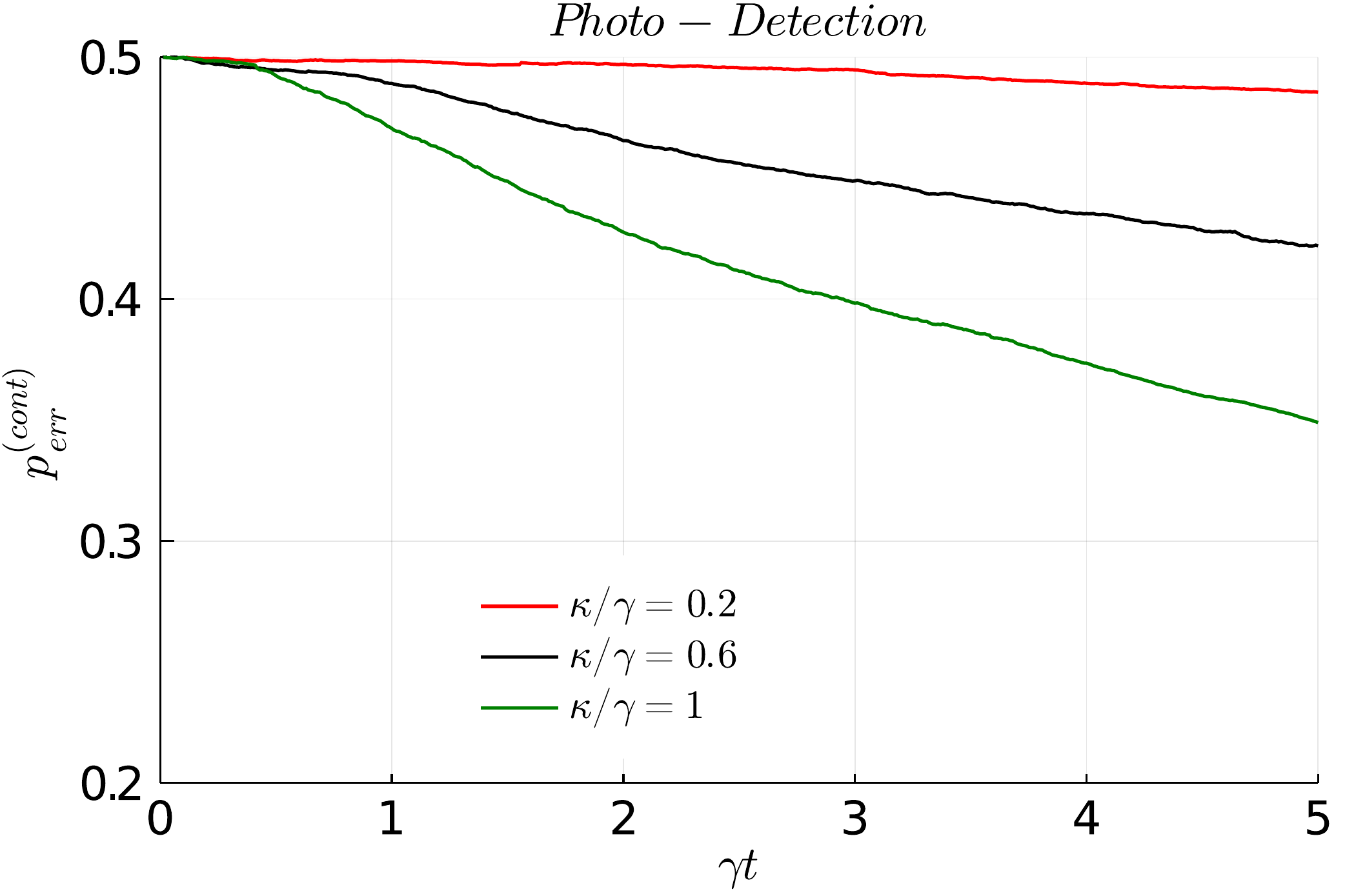}
\put(20,45){$\hat{c}=\hat{\sigma}_-$}
\put(90,65){(a)}
\end{overpic}
\begin{overpic}[width=.3\linewidth]{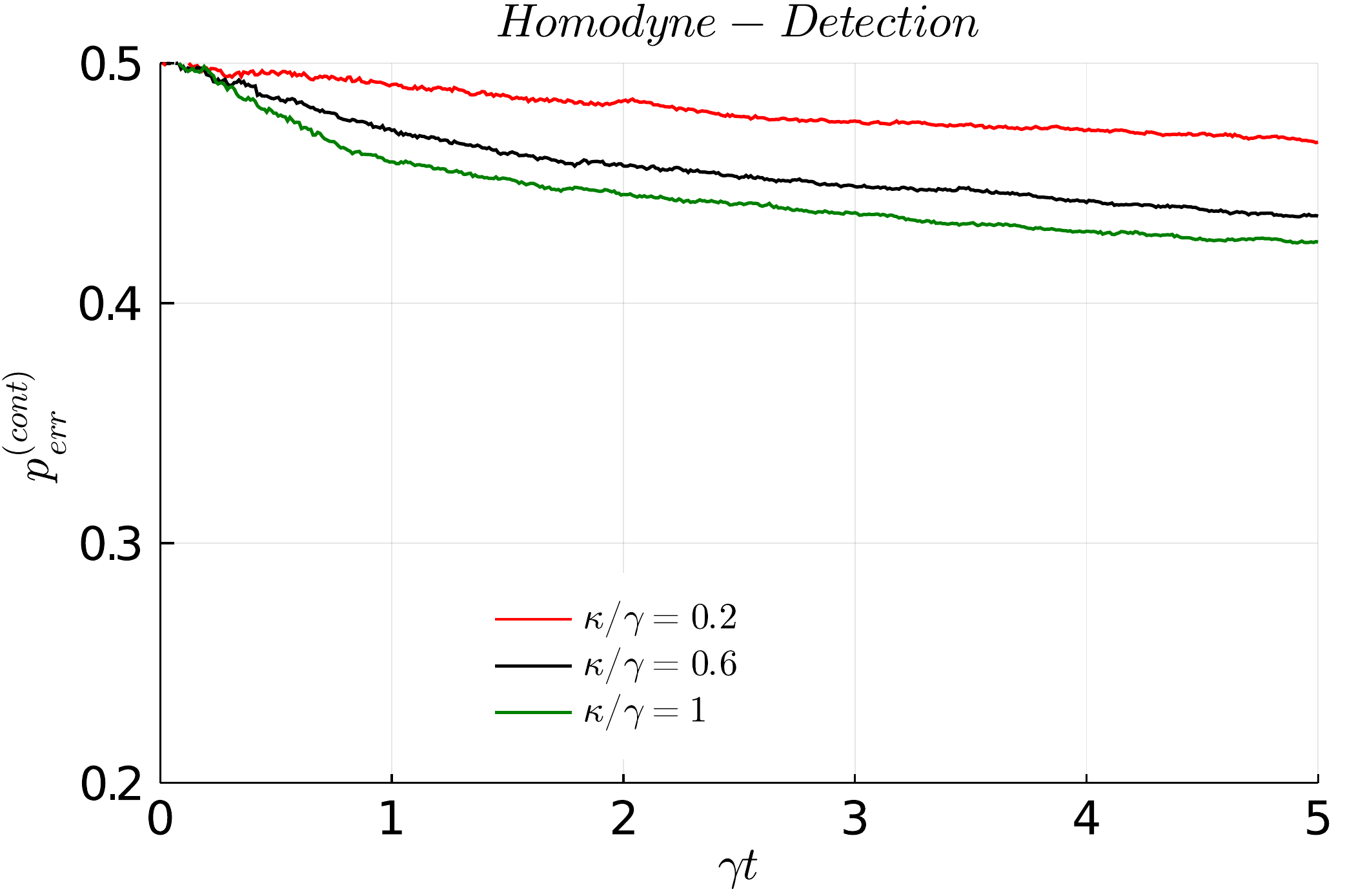}
\put(20,45){$\hat{c}=\hat{\sigma}_-$}
\put(90,65){(b)}
\end{overpic}
\begin{overpic}[width=.3\linewidth]{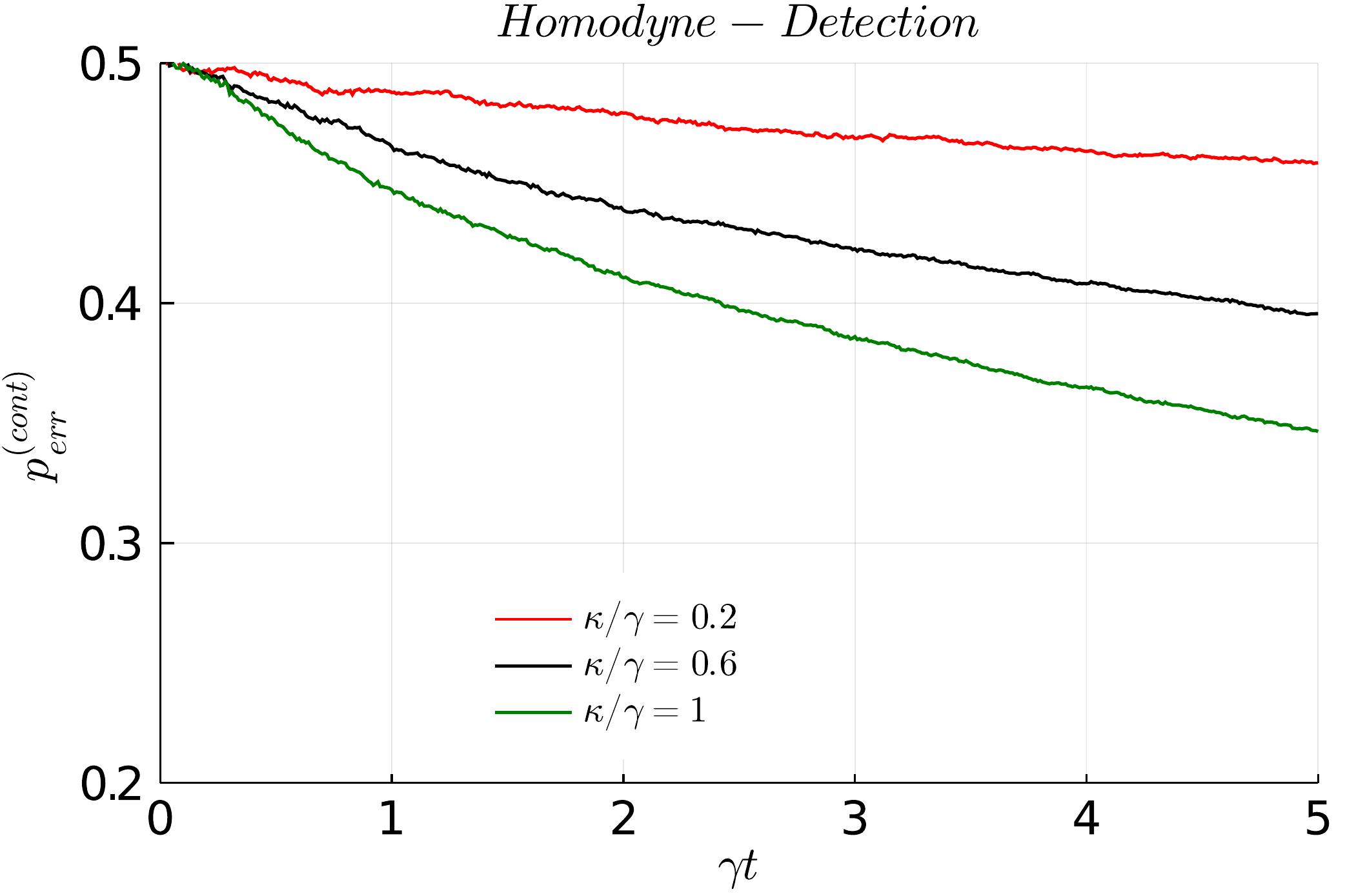}
\put(20,35){$\hat{c}=\hat{\sigma}_x/2$}
\put(90,65){(c)}
\end{overpic}
\\
\vspace{.5cm}
\begin{overpic}[width=.3\linewidth]{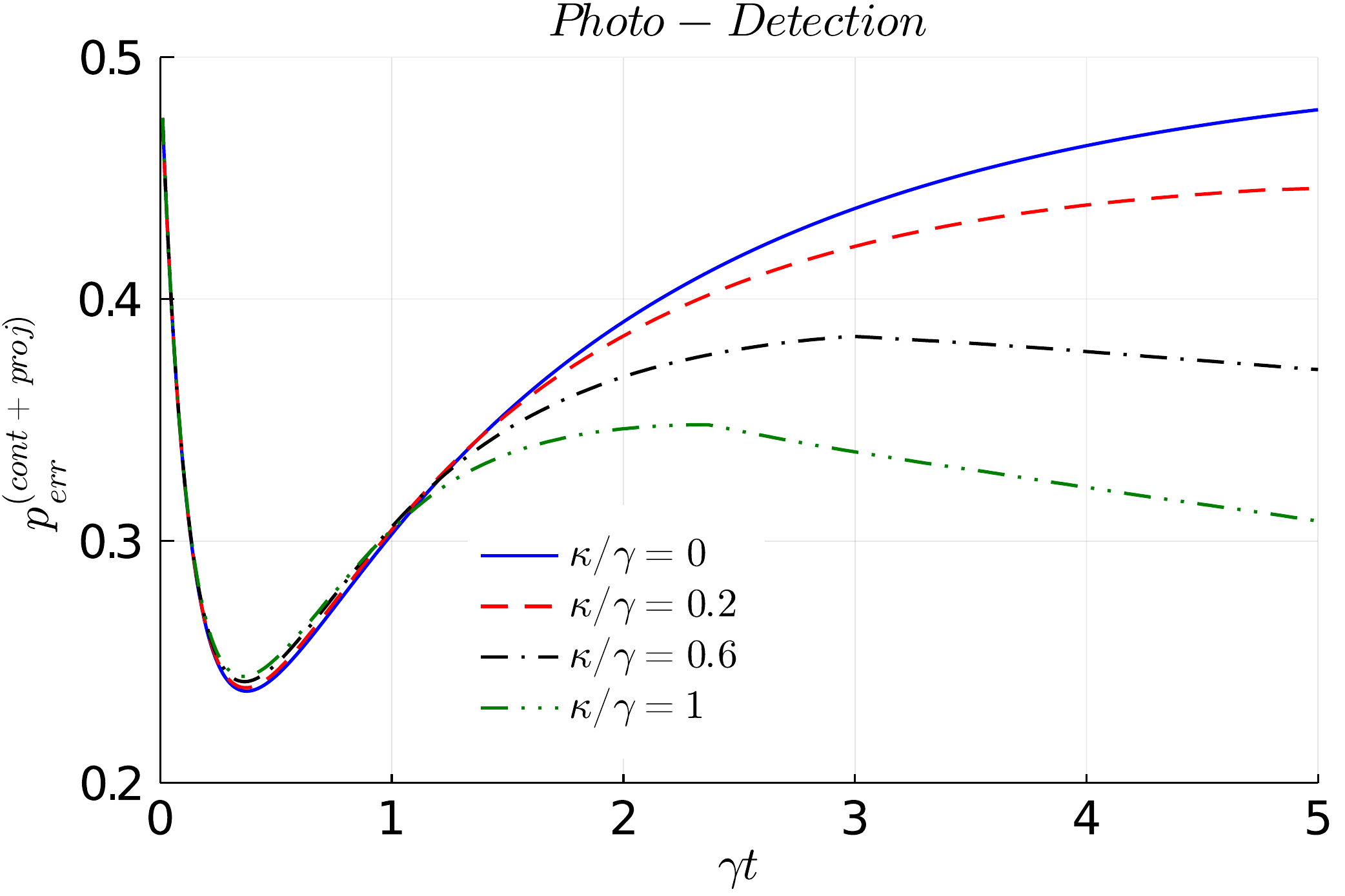}
\put(20,45){$\hat{c}=\hat{\sigma}_-$}
\put(90,65){(d)}
\end{overpic}
\begin{overpic}[width=.3\linewidth]{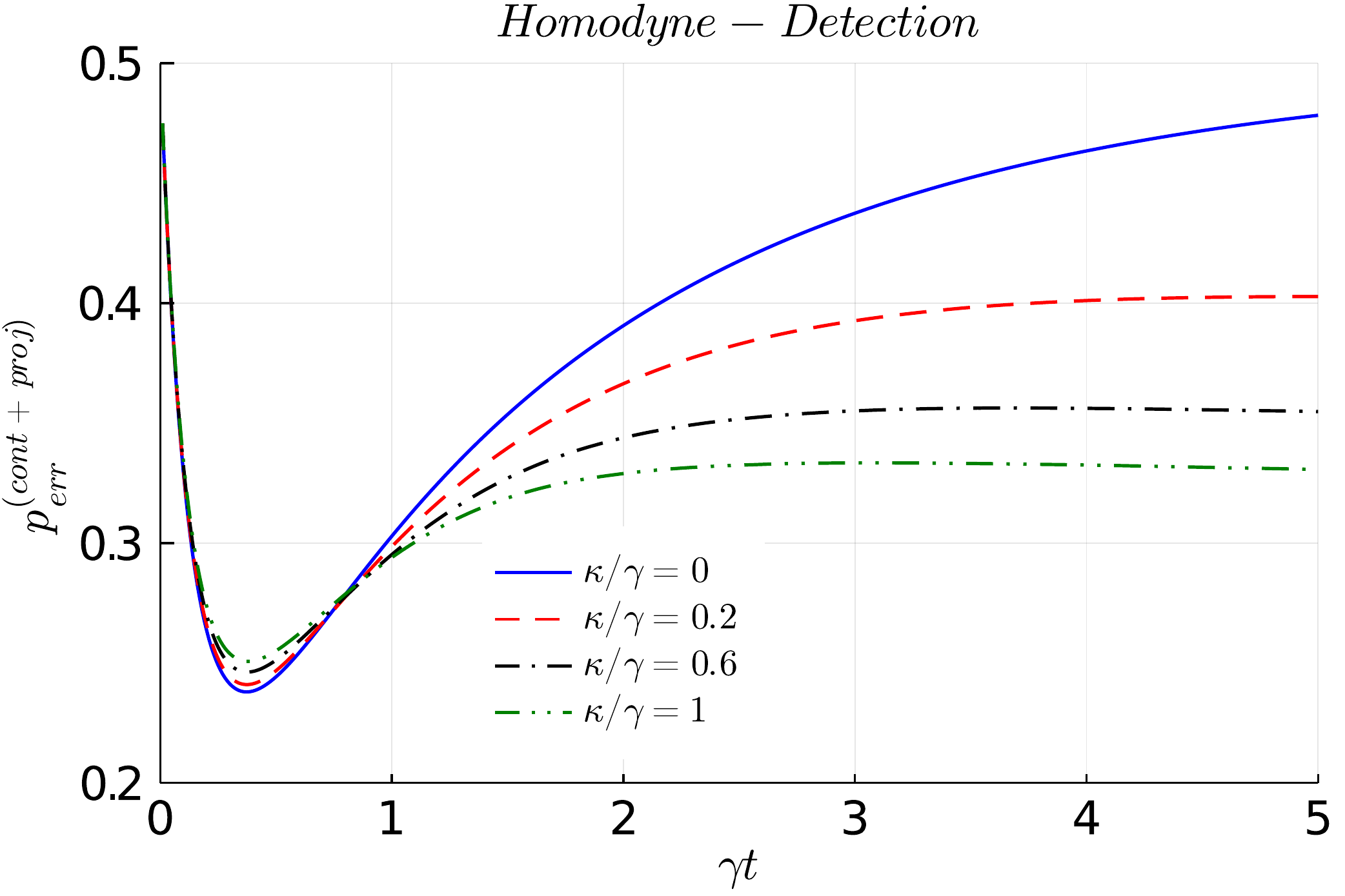}
\put(20,45){$\hat{c}=\hat{\sigma}_-$}
\put(90,65){(e)}
\end{overpic}
\begin{overpic}[width=.3\linewidth]{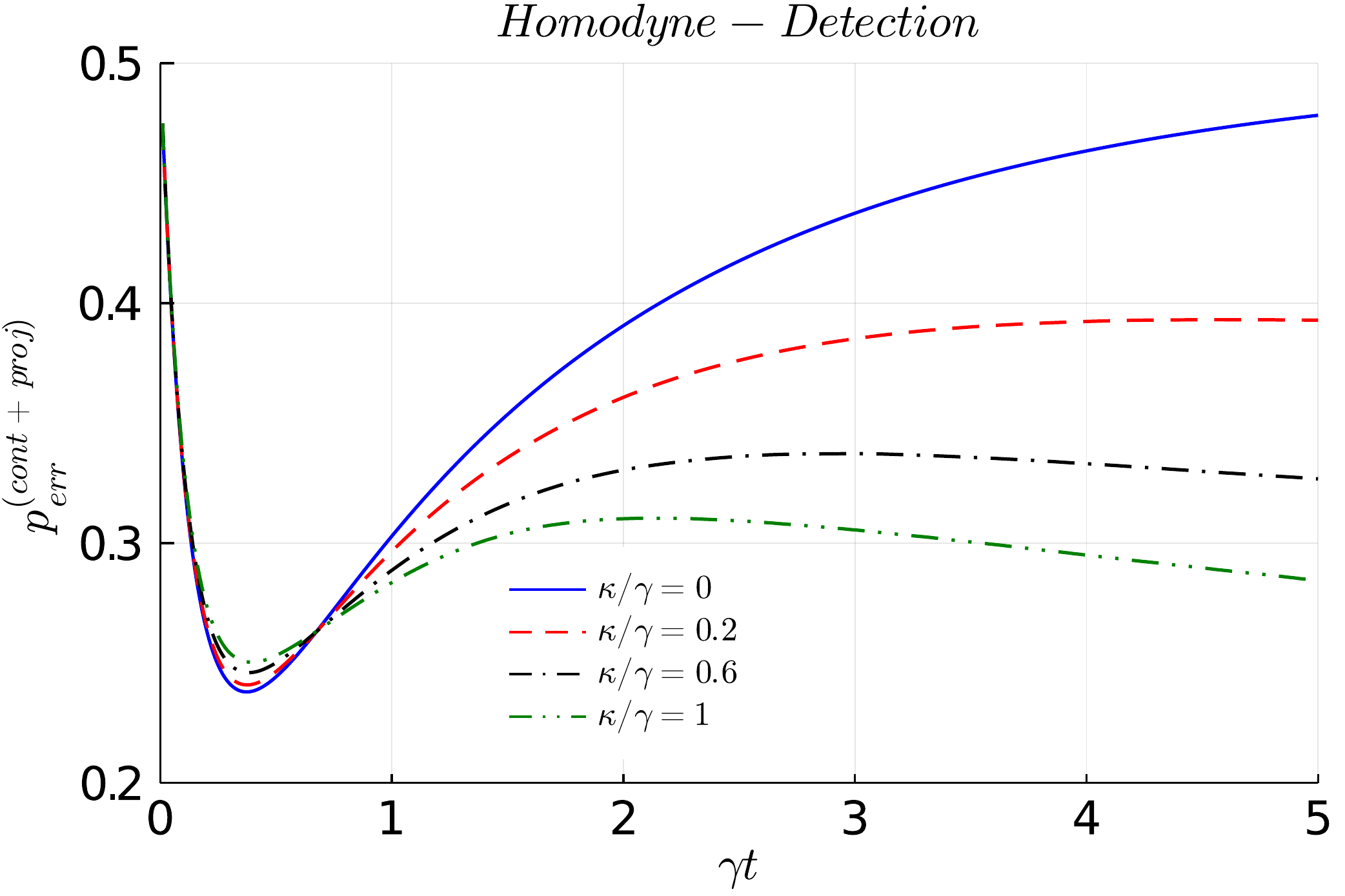}
\put(20,45){$\hat{c}=\hat{\sigma}_x/2$}
\put(90,65){(f)}
\end{overpic}
\caption{\label{fig:comparing-setups-GHZ}
Setting 
$
\ket{\Phi^+}
$
as initial state of the $SA_1$ system, we plot (a)-(c) $p_{err}^{(cont)}(t)$ and (d)-(f) $p_{err}^{(cont + proj)}(t)$ for different detection strategies:
for $\hat{c}=\hat{\sigma}_-$
Photo-detection (a),(d) and 
Homodyne-detection (b),(e); 
and 
for $\hat{c}=\hat{\sigma}_x/2$
Homodyne-detection (c),(f).
Different curves refer to different values of $\kappa$ as indicated in the legend with $\kappa=0$ referring to the case where $A$ is decoupled from the probe. 
All data in this figure are obtained 
in the isothermal QBT scenario  $\beta_B=\beta_F=\beta$,
setting
$\beta \omega_0=1/5.5$ and $\eta=1$.
}
\end{figure*}

\begin{figure}
\vspace{1cm}
\centering

\begin{overpic}[width=.98\linewidth]{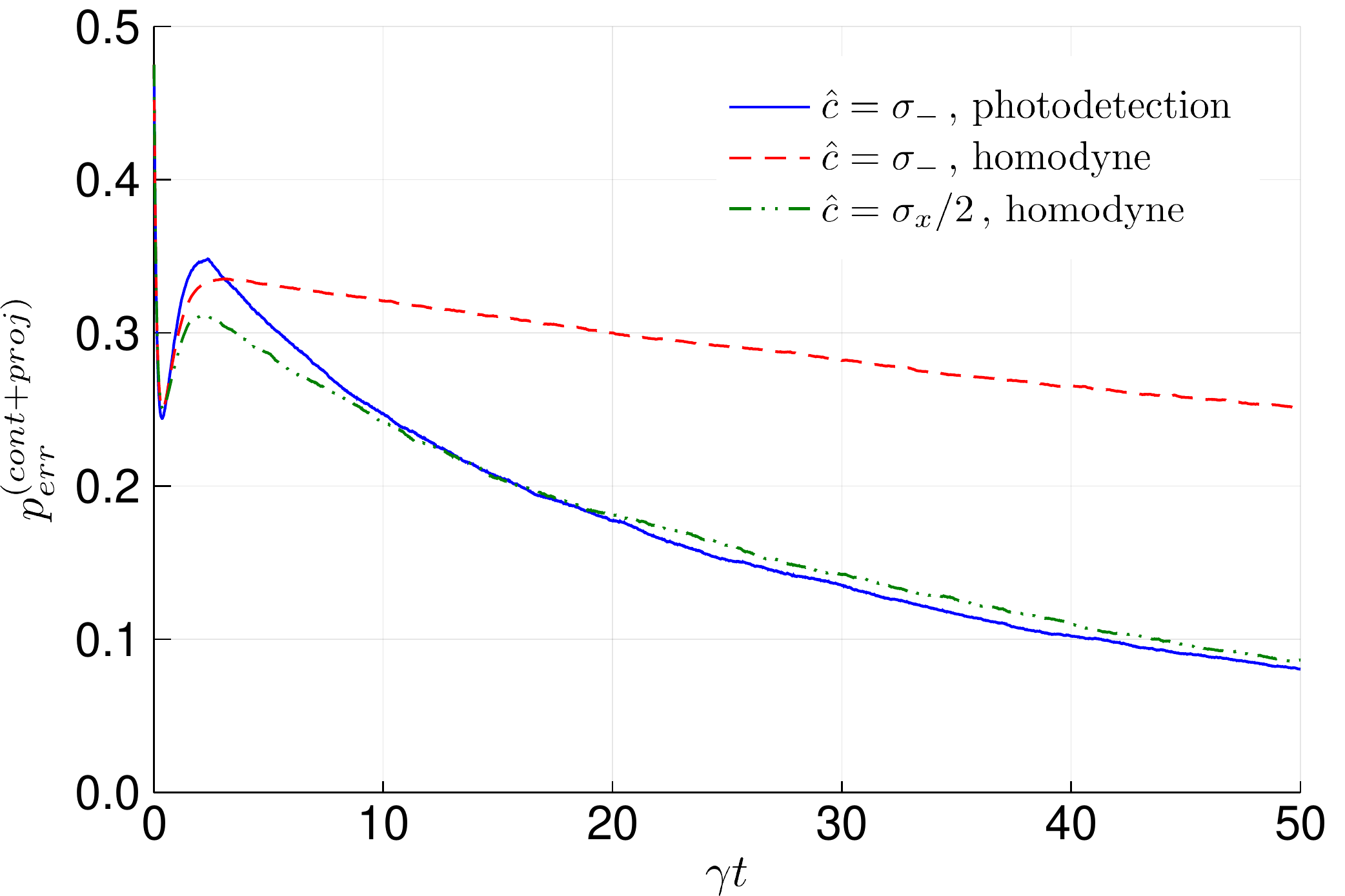}
\end{overpic}

\caption{\label{fig:comparison}
Long time behaviour of $p_{err}^{(cont+proj)}(t)$ for different continuous monitoring  strategies
(see legend). 
All data are obtained 
in the isothermal QBT scenario  $\beta_B=\beta_F=\beta$,
setting
$\beta \omega_0=1/5.5$, $\eta=1$, $\kappa/\gamma=1$ and considering $\ket{\Phi^+}$ as the initial state.
}
\end{figure}

We then fix the monitoring efficiency to its maximum value $\eta=1$ and turn our attention to the coupling 
$\kappa$ that gauges the $S$-$A$ coupling, which in this framework can be also interpreted as a measurement strength. The results are depicted in the panels (a), (b), (d) and (e) of Fig.~\ref{fig:comparing-setups-GHZ} where in the top (bottom) panels we show the behaviour of $p_{err}^{(cont)}(t)$ ($p_{err}^{(cont+proj)}(t)$) for different values of $\kappa$. The first thing one may notice is that 
for  low values of $\kappa$ photo-detection is less efficient than homodyne in reducing $p_{err}^{(cont)}(t)$, while for large values of $\kappa$ it becomes the preferable choice -- see panels (a) and (b).
Regarding $p_{err}^{(cont+proj)}(t)$ independently on the type of detection on $A$, we can make two relevant observations: (i) at short time scales the monitoring of $A$ does not lead to a better discrimination as indeed the optimal value still corresponds to the case $\kappa=0$ (blue curves in the figure); (ii) on the other hand, at long time scales the cumulative information acquired by continuous monitoring definitely improves discrimination for increasing values of $\kappa$. In particular we have numerical evidence that both $p_{err}^{(cont)}(t)$ and $p_{err}^{(cont+proj)}(t)$ go to zero in the long time limit, and thus that in general the minimum error probability obtainable for $\kappa=0$ can be overcome by considering either $\kappa$ and/or time large enough -- see Fig.~\ref{fig:comparison}. 
\subsubsection{Dispersive $S$-$A$ coupling} 

Consider next the possibility of coupling dispersively the system to the environment $A$ represented by taking  $\hat{c} = \hat{\sigma}_x$ as the jump operator of the model~\cite{hacohen-gourgy_quantum_2016}. We start by observing that, as $\hat{c}^\dag \hat{c} = \hat{\sigma}_x^2 = \mathbbm{1}$, the probability for continuous photodetection is independent on the state, and thus it cannot contain any information on the bath~$E$. For this reason for the photo-detection unravelling one would obtain $p_{err}^{(cont)}(t)=1/2$ at any time~$t$. We thus show the result of $p_{err}^{(cont)}(t)$ for homodyne detection only, see panel (c) of Fig.~\ref{fig:comparing-setups-GHZ}, with
the corresponding case of final projective measurement on $SA_1$ in panel (f). Also in this case, we find that at short time scales the coupling with $A$ and the monitoring is not helpful, as the best performances are observed for $\kappa=0$. On the other hand we find that $p_{err}^{(cont+proj)}(t)$ decreases towards zero at long time scales and that, as in the previous case, better results are obtained by increasing the coupling $\kappa$. We do not provide results for continuous photodetection with a final projective measurement as, while error probabilities below $1/2$ are observed, the performances are definitely worse respect to the other  cases we have considered. 
\\

\subsubsection{Strategies Comparision}
We compare the three different strategies, continuous homodyne and photodetection with $\hat{c}=\hat{\sigma}_-$ and continuous homodyne with $\hat{c}=\hat{\sigma}_x/2$ in Fig. \ref{fig:comparison}. We observe that in the long-time limit the two best strategies correspond to either performing continuous homodyne on an environment coupled dispersively via the jump operator $\hat{c}=\hat{\sigma}_x/2$ or continuous photodetection with jump operator $\hat{c}=\hat{\sigma}_-$. Moreover the first strategy is also the best one in the short-time limit (we remark that similar results are obtained numerically for different values of the parameters).

\section{Conclusions and final remarks} 
\label{sec-experimental-realizations}

In this work we have  investigated the possibility of
improving  the performances of QBT task originally presented in~\cite{farina2019tagging,gianani2020discrimination} using extra auxiliary resources 
such as an extra memory element $A_1$ that could be initially entangled with the original probe $S$, and an extra environment $A$ that instead is allowed to interact with the $S$ while possibly being monitored continuously in time. 
In particular we notice that  the QBT task can benefit even when   $A$ is monitored very inefficiently ($\eta=0$), an effect that for instance is observed  in the equal temperature case which for 
long interaction time would not allow for QBT discrimination in the  original proposal~\cite{farina2019tagging,gianani2020discrimination}.
We finally compared 
 the performances associated to different realizations of continuous monitoring of $A$
via photo-detection or homodyne,  proving 
that a finite detection efficiency is naturally beneficial to the QBT task. 
Before concluding we would like to comment that the reported results, while derived in the specific QBT setting of ~\cite{farina2019tagging,gianani2020discrimination} can be generalized to improve the performances of arbitrary quantum hypothesis tasks, in particular in all 
those problems where an agent is asked to use an external probe to discriminate between alternative quantum trajectories associated with different dynamical quantum generators. 
We also would like to mention that experimental realizations  for specific setup we have analyzed in the manuscript are feasible e.g. in the context of  superconducting qubits \cite{hacohen-gourgy_quantum_2016,ficheux2018dynamics}. In these models, assuming  $S$ and $A_1$ to be 
superconducting transmon qubits, the initial entanglement configuration between them can be reached for example with the use of a common bus resonator \cite{egger2018entanglement}. 
Notice also  that configurations where $S$ is capacitively coupled with two baths  $E$ and $A$ (the latter being continuously monitored)  are now experimentally under control, e.g. interpreting the $S$ as a quantum valve~\cite{pekola2018valve}. 
In particular in our case the engineered environment $A$ may consist in a cavity where this time only $S$  is embedded and the initial coupling with $A_1$ is now off-detuned, and  the transmission of  input microwave fields are used for quadrature and dispersive measurements~\cite{tan2015prediction}. Specifically either a fluorescence measurement \cite{ficheux2018dynamics}, corresponding to a jump operator $\hat{c}=\hat{\sigma}_-$, or a dispersive measurement \cite{hacohen-gourgy_quantum_2016}, corresponding for example to a jump operator $\hat{c}=\hat{\sigma}_x/2$, is performed by using a resonant field. The output for homodyne measurement is instead recorded via a Josephson parametric amplifier, or, in the case of heterodyne measurements, via a Josephson parametric converter \cite{ficheux2018dynamics}.
The final Helstrom measurement on $SA_1$ can be generally achieved in these experiments by applying a strong, dispersively coupled probe field \cite{murch2013observing,tan2015prediction, kiilerich2018hypothesis}.
\\

 V.G. acknowledges MIUR  (Ministero dell' istruzione, dell' Universita' e della Ricerca) via project PRIN 2017 ``Taming complexity via QUantum Strategies a Hybrid Integrated Photonic approach" (QUSHIP) Id. 2017SRNBRK.

\bibliography{bibfile}

\begin{widetext}
\appendix

\section{Numerical integration of stochastic master equations}
\label{a:numerical}
We here describe the method proposed in \cite{rouchon2014models,rouchon2015efficient} in order to efficiently numerically integrate SMEs, as the ones reported in Eqs.~\eqref{eq:SMEpd} and \eqref{eq:SMEhd}. One proves that the quantum state solution of these SMEs can be written after each time step $dt$ as
\begin{align}
    {\hat{\varrho}}^{c}_{q}(t+dt) = \frac{\sum_k \hat{M}^{(k)}_{x_t}\left[{\hat{\varrho}}^{c}_{q}(t)+ \mathcal{L}_{q} {\hat{\varrho}}^{c}_{q}(t) dt \right]{\hat{M}_{x_t}^{(k)}}{}^\dag}{\Tr(\sum_k \hat{M}^{(k)}_{x_t}\left[{\hat{\varrho}}^{c}_{q}(t)+ \mathcal{L}_{q} {\hat{\varrho}}^{c}_{q}(t) dt \right]{\hat{M}_{x_t}^{(k)}}{}^\dag)}\,,
    \label{eq:rouchonralph}
\end{align}
where we have introduced the Kraus operators $\hat{M}_{x_t}^{(k)}$ that describe the effect of the measurement, with outcome $x_t$, on the quantum state at each time $t$. The form of these operators depends on the kind of measurement that is performed. In the case of photodetection, the two Kraus operators corresponding to the two possible measurement outcomes $x_t=\{0,1\}$ are
\begin{align}
\begin{cases}
     \hat{M}_0^{(1)} = \hat{\mathbbm{1}} - \frac{\kappa}{2} \hat{c}^\dag \hat{c} \,dt\\
     \hat{M}_0^{(2)} = \sqrt{(1-\eta) \kappa dt}\, \hat{c}
\end{cases}
    \,, \qquad \qquad 
    \hat{M}_1^{(1)} = \sqrt{\eta \kappa dt} \, \hat{c} \,,
\end{align}
that are applied according to the Poisson increment probabilities $p_0 = 1 - \eta \kappa \Tr[{\hat{\varrho}}^{c}_{q} \hat{c}^\dag \hat{c} ] dt$ and $p_1 =\eta \kappa \Tr[{\hat{\varrho}}^{c}_{q} \hat{c}^\dag \hat{c} ] dt$. As regards continuous homodyne detection, the continuous outcome corresponds to the photocurrent $x_t = dy_t$ and the corresponding Kraus operators have the form
\begin{align}
\begin{cases}
\hat{M}_{dy_t}^{(1)} =\hat{\mathbbm{1}} - \frac{\kappa}{2} \hat{c}^\dag \hat{c} \,dt + \sqrt{\eta \kappa} \hat{c} \, dy_t\\
\hat{M}_{dy_t}^{(2)} = \sqrt{(1-\eta) \kappa dt}\, \hat{c}
\end{cases}
     \,,
\end{align}
where the randomness of the process is originated by the Wiener increment entering in the formula for the photocurrent~(\ref{eq:photocurrent}). 
\par
This numerical method also allows to evaluate straightforwardly the likelihood of each trajectory. In fact, at each time step, the likelihood of obtaining the measurement outcome $x_t$ can be evaluated by taking the trace of the operator at the numerator in Eq. (\ref{eq:rouchonralph}), i.e.
\begin{align}
 l_{x_t} = \Tr[\tilde{\varrho}_q^c(t+dt)] \,,
\end{align}
where
\begin{align}
  \tilde{\varrho}_q^c(t+dt) &= \sum_k \hat{M}^{(k)}_{x_t}\left[{\hat{\varrho}}^{c}_{q}(t)+ \mathcal{L}_{q} {\hat{\varrho}}^{c}_{q}(t) dt \right]{\hat{M}_{x_t}^{(k)}}{}^\dag \,.
\end{align}
As remarked in the main text, by assuming to start and stop the  monitoring respectively at time $t_0$ and time $t$, each trajectory can be identified by the string of records $D_t$ of Eq.~(\ref{eq:string1}):
the corresponding likelihood can thus be evaluated as
\begin{align}
    L(D_t|q) = \prod_{t^\prime = t_0}^t l_{x_{t^\prime}} = \prod_{t^\prime = t_0}^t \Tr[\tilde{\varrho}_q^c(t^\prime+dt)]\;.  \label{eq:likelihood}
\end{align}

\section{Steady state for a multichannel master equation}
\label{app:steady}

When we are not continuously monitoring the bath $A$, the dynamical evolution of $S$ is described by 
the master equation~(\ref{eq:ME}) whose dynamical generator is given by the super-operator 
\begin{eqnarray}
\label{generator-MC}
 \mathcal{L}^{(ext)}_{q}\bullet :=\mathcal{L}_{q}\bullet +  \kappa \mathcal{D}_{[\hat{c}]}\bullet = 
 -i \Big[\hat{H}_S,\bullet \Big]
+\gamma_q^{-} \nonumber
\mathcal{D}_{[\hat{\sigma}_-]}\bullet
+
\gamma_q^{+}
\mathcal{D}_{[\hat{\sigma}_+]}\bullet  
+ \gamma_q^{x}
\mathcal{D}_{[\frac{\hat{\sigma}_x}{2}]}\bullet \;,  
\end{eqnarray}
where for
two cases considered in the main text $\hat{c} =\{ \hat{\sigma}_-, \hat{\sigma}_x/2\}$ we have
\begin{align} 
  {\rm For} \, \, \hat{c} &= \hat{\sigma}_- :\quad \quad \gamma^-_q = \gamma (1+ s_q N_q(\beta_q)) + \kappa; \;   \quad \gamma^+_q = \gamma N_q(\beta_q); \;  \quad \gamma^x_q = 0 \nonumber, \\
  {\rm For}  \,\, \hat{c} &= \frac{\hat{\sigma}_x}{2} :\quad \quad  \; \gamma^-_q = \gamma (1+ s_q N_q(\beta_q)); \;  \quad \quad \;\:\: \gamma^+_q = \gamma N_q(\beta_q); \; \quad \gamma^x_q = \kappa. \label{3cond}
\end{align}
%
To discuss the statistics tagging in the long time limit  we solve the equation  $\mathcal{L}^{(ext)}_{q}{\hat{\varrho}}_{q}^{ss} =0$ which, irrespectively from the input state of the system, provides the  steady state~${\hat{\varrho}}_{q}^{ss}$ solution of the 
system dynamics, i.e. 
\begin{eqnarray} 
\lim_{t\rightarrow \infty} {\hat{\varrho}}_{q}(t)  =  {\hat{\varrho}}_{q}^{ss}\;. 
\end{eqnarray} 
Writing hence ${\hat{\varrho}}_{q}^{ss} = p_q \ket{1}\bra{1} + (1-p_q) \ket{0}\bra{0} + c_q \ket{0}\bra{1} + c^*_q \ket{1}\bra{0} $ we obtain the following conditions
\begin{equation} \label{condpop} p_q = \frac{\gamma_q^+ + \gamma_q^x/4}{\gamma_q^+ +\gamma_q^- + \gamma_q^x/2}  ,\quad \qquad   c_q= 0\;. \end{equation} 
Notice that irrespectively from the selected QBT hypothesis the off-diagonal elements are always null. On the other hand 
the associated conditions for the populations at steady state are obtained by plugging \eqref{3cond} in the equation \eqref{condpop}:
\begin{align} \nonumber
  {\rm For} \, \, \hat{c} &= \hat{\sigma}_- :\quad  \quad   p_B =  \frac{\gamma N_B(\beta_B)}{\gamma (2 N_B(\beta_B)+1) +\kappa }   , \quad p_F = \frac{\gamma N_F(\beta_F)}{\gamma +\kappa },   \\ 
  {\rm For}  \,\, \hat{c} &= \frac{\hat{\sigma}_x}{2} :\quad\quad   p_B =  \frac{\gamma N_B(\beta_B) + \kappa/4}{\gamma (2 N_B(\beta_B)+1) +\kappa/2 }   , \quad p_F = \frac{\gamma N_F(\beta_F) + \kappa/4}{\gamma +\kappa/2 } . \label{steadyp}
\end{align}
With the above expressions we can now express the asymptotic limit of the HEP functional~(\ref{Helstrom}) 
\begin{eqnarray}\label{HEPBOUNDasymp} 
\lim_{t\rightarrow \infty} {\rm HEP}(t;{\hat{\varrho}}(0)) = 
{\rm HEP}(t\rightarrow \infty)  &:=& \frac{1}{2} \left(1- \frac{ \|  {\hat{\varrho}}_B^{ss}- {\hat{\varrho}}_F^{ss}\|_1}{2}\right)=\frac{1}{2} \left(1- |p_B-p_F  |  \right) .  
\end{eqnarray} 
We can also represent the figure of merit in terms of the heat flowing between the two environments at steady state.
The heat flowing in $A$ is characterized in terms of the following 
equation~\cite{alicki2018introduction}
\begin{equation} 
\dot{Q}^{(E_{q} \Rightarrow A) } =- \kappa {\rm Tr}[\hat{H}_S \mathcal{D}_{[\hat{c}]} \hat{\varrho}^{ss}_q], \label{heat}
\end{equation}
that in the case of $\hat{c}=\hat{\sigma}_-$ gives $\omega_0 \kappa p_{q}$.
 Combining equation \eqref{heat} 
with the first of Eqs. \eqref{steadyp} it is straightforward to obtain the results \eqref{hel-0eta} and \eqref{flow} of the main text, after expressing the Fermi function in terms of the Bose function for uniforming the notation $N_F(\beta_F) = \frac{N_B(\beta_F)}{1+ 2 N_B(\beta_F)} $.
Also in the case of $\hat{c}= \hat{\sigma}_x/2$ we are able to establish a connection between the modulus of the population difference and the heat flow.
Using the definition \eqref{heat} we have
$
- \kappa {\rm Tr} \big[\hat{H}_S \mathcal{D}_{[\frac{\hat{\sigma}_x}{2}]}{\hat{\varrho}}_{q}^{ss} \big]  = - \frac{\omega_0 \kappa }{4}(1- 2p_q) $
from which we derive 
\begin{equation}
|p_B - p_F| = \frac{2}{\omega_0 \kappa} | \dot{Q}^{(E_{B} \Rightarrow A) } - \dot{Q}^{(E_{F} \Rightarrow A) }  |,  
\end{equation}
and hence 
\begin{eqnarray}\label{HEPBOUNDasympC}  
{\rm HEP}(t\rightarrow \infty)  &=& \frac{1}{2} \left(1- \frac{2}{\omega_0 \kappa} \left| \dot{Q}^{(E_{B} \Rightarrow A) } - \dot{Q}^{(E_{F} \Rightarrow A) }  \right|\right) .  
\end{eqnarray}
At last, we discuss how the tagging procedure can be influenced by the coupling with the bath $A$, gauged through the parameter $\kappa$.
Choosing the value of $\kappa$ for which Eq. \eqref{hel-0eta} is equal to $1/2$ allows to find $\kappa_c$ in Eq. \eqref{critical}.
The same analysis for the case with $\hat{c}=\frac{\hat{\sigma}_x}{2}$ leads to the following critical value
\begin{equation}
\frac{\kappa_c}{\gamma} = 2     \frac{N_B(\beta_B)}{N_F(\beta_F)} -1.
\end{equation}
The optimal value for the discrimination at steady state when $\beta_F = \beta_B = \beta $, instead, is obtained by optimizing the figures of merit with respect to $\kappa$.
In this way the results \eqref{Kbest} and 
\begin{equation}
    \frac{\kappa_{best}}{\gamma} = 2 \sqrt{2 N_B(\beta) +1}.
\end{equation}
are found for $\hat{c}= \hat{\sigma}_-$ and $\hat{c}= \hat{\sigma}_x/2$, respectively.

\end{widetext}

\end{document}